\documentclass[aps,10pt,pra,twocolumn,amsmath,amssymb,superscriptaddress,groupedaddress]{revtex4-1}

\usepackage[english]{babel} 
\usepackage[utf8]{inputenc}
\usepackage[T1]{fontenc}
\usepackage{graphicx}
\usepackage{epsfig}
\usepackage{graphicx}
\usepackage{subfigure}

\usepackage{hyperref}
\hypersetup{
	colorlinks=true,
	linkcolor=red,			
	citecolor=green,		
	filecolor=magenta,		
	urlcolor=magenta		
}

\newcommand{\ket}[1]{\left\vert #1 \right\rangle}
\newcommand{\bra}[1]{\left\langle #1 \right\vert}
\newcommand{\abs}[1]{\left\vert #1 \right\vert}
\newcommand{\norm}[1]{\vert \vert #1 \vert \vert}

\newcommand{\komm}[2]{\left[ #1 , #2 \right]}
\newcommand{\akomm}[2]{\left\{ #1 , #2 \right\}}
\newcommand{\arccot}{\operatorname{arccot}}
\renewcommand{\d}{\mathrm{d}}
\newcommand{\comma}{~,}
\newcommand{\fullstop}{~.}
\newcommand{\Tr}{\mathrm{Tr}}

\begin{document}

\selectlanguage{english}

\title{Optimal synchronization deep in the quantum regime: resource and fundamental limit}

\author{Martin Koppenh\"ofer}
\affiliation{Department of Physics, University of Basel, Klingelbergstrasse 82, CH-4056 Basel, Switzerland}

\author{Alexandre Roulet}
\affiliation{Department of Physics, University of Basel, Klingelbergstrasse 82, CH-4056 Basel, Switzerland}

\date{\today}

\begin{abstract}
	We develop an analytical framework to study the synchronization of a quantum self-sustained oscillator to an external signal. 
	Our unified description allows us to identify the resource on which quantum synchronization relies, and to compare quantitatively the synchronization behavior of different limit cycles and signals. 
	We focus on the most elementary quantum system that is able to host a self-sustained oscillation, namely a single spin 1. 
	Despite the spin having no classical analogue, we first show that it can realize the van der Pol limit cycle deep in the quantum regime, which allows us to provide an analytical understanding to recently reported numerical results. 
	Moving on to the equatorial limit cycle, we then reveal the existence of an interference-based quantum synchronization blockade and extend the classical Arnold tongue to a snake-like split tongue. 
	Finally, we derive the maximum synchronization that can be achieved in the spin-$1$ system, and construct a limit cycle that reaches this fundamental limit asymptotically.
\end{abstract}

\maketitle

\def\ValueFigureItheta{1.5}
\def\ValueFigureIphi{1}
\def\ValueFigureIgammag{1}
\def\ValueFigureIgammad{100}

\def\ValueFigureIIDelta{0}
\def\ValueFigureIIgammag{1}
\def\ValueFigureIIeta{0.1}
\def\ValueFigureIIgammadimb{10}
\def\ValueFigureIIgammadbal{1}
\def\ValueFigureIItop{0.5}
\def\ValueFigureIItmo{0.5}
\def\ValueFigureIItmp{0}
\def\ValueFigureIIepsA{my\[Epsilon]1}
\def\ValueFigureIIepsB{3}
\def\ValueFigureIIepsC{100}
\def\ValueFigureIIepsThresBal{0.0707107}
\def\ValueFigureIIepsThersImb{0.0995037}

\def\ValueFigureIIIgammag{1}
\def\ValueFigureIIIgammad{100}
\def\ValueFigureIIIeta{0.1}
\def\ValueFigureIIItop{0.5}
\def\ValueFigureIIItmo{0.5}
\def\ValueFigureIIItmp{0}

\def\ValueFigureIVgammag{1}
\def\ValueFigureIVgammad{1000}
\def\ValueFigureIVeta{0.1}

\def\ValueFigureIVbgammag{1}
\def\ValueFigureIVbgammad{100}
\def\ValueFigureIVbgammadB{1000}
\def\ValueFigureIVbgammadC{10}
\def\ValueFigureIVbeta{0.1}

\def\ValueFigureVgammag{1}
\def\ValueFigureVgammad{1}
\def\ValueFigureVeta{0.1}
\def\ValueFigureVDelta{0}

\def\ValueFigureVIgammag{1}
\def\ValueFigureVIgammadA{1}
\def\ValueFigureVIgammadB{100}
\def\ValueFigureVIgammadC{10000}
\def\ValueFigureVIeta{0.1}
\def\ValueFigureVIchi{0}

\def\ValueFigureVIIIgammag{1}
\def\ValueFigureVIIIgammad{100}
\def\ValueFigureVIIIDelta{0}
\def\ValueFigureVIIIrA{0.5}
\def\ValueFigureVIIIrB{2.5}
\def\ValueFigureVIIIrC{4}
\def\ValueFigureVIIIrD{9}

\section{Introduction}
Since the first observation reported by Huygens four centuries ago~\cite{Huygens1673}, synchronization~\cite{Pikovsky-Synchronization} has provided a universal framework to capture features shared by very different complex systems, such as chaotic electronic circuits and biological neuron networks~\cite{pecora90,Chagnac-JNeurophysiol.62.1149,Ferrari-NeuralNetworks.66.107,rodrigues16}. 
The essence of synchronization is the ability of a self-sustained oscillator to adjust its rhythm when subjected to a weak perturbation.

Recently, significant progress has been made in understanding whether quantum systems could synchronize as well. 
In particular, the van der Pol oscillator, a classic self-sustained oscillator extensively used in biology~\cite{vdp28,rowat93,kronauer98,rompala07}, has been investigated in the quantum regime of a few excitations~\cite{Lee-PRL.111.234101,Walter-PRL.112.094102}, demonstrating that synchronization to a semi-classical signal survives in this limit despite the inevitable presence of quantum noise. 
Since then, this system has been used to probe the features of quantum synchronization~\cite{talitha17}, such as the role of the number-phase uncertainty~\cite{armour18} or the exciting possibility to enhance synchronization by applying a squeezing signal~\cite{Sonar-PRL.120.163601}. 
Yet, the infinite-dimensional Hilbert space combined with the intrinsic non-linear and dissipative dynamics have limited studies to numerical explorations of the parameter space, usually guided by an analytical description of the classical limit.

Addressing this challenge of understanding quantum synchronization beyond numerics, an elementary unit -- a spin 1 -- has recently been identified as the smallest quantum system that can be synchronized~\cite{Roulet-PRL.121.053601}. 
Its finite Hilbert space of dimension 3 has already proved useful to clarify analytically the relation between entanglement and quantum synchronization~\cite{Roulet-PRL.121.063601}. 
Here, we consider a spin $1$ subjected to an external signal and aim to analytically understand the resources on which quantum synchronization relies, the role of quantum effects, and by which means synchronization can reach the fundamental limit imposed by the laws of quantum mechanics.

To put the spin-$1$ platform on solid grounds, the first question we address is whether this minimal system with no classical analogue is actually complex enough to capture all the features of quantum synchronization that appear in classically-inspired systems like the van der Pol oscillator. 
We answer this question by bridging the gap between the two architectures, demonstrating that a van der Pol oscillator operating deep in the quantum regime can be represented in the spin 1 platform, even though the spin phase space lives on a sphere and does not correspond to a position-momentum representation.
This result allows us to connect with previous numerical findings obtained on a  harmonic-oscillator platform, and to further improve on them thanks to the analytical accessibility of the spin-$1$ system. 
In particular, we identify the coherences between energy levels as the resource for quantum synchronization and we find that while squeezing does improve the phase locking of a van der Pol limit cycle, an even better performance can be achieved by additionally modifying the semi-classical component of the signal. 
We prove that this signal yields the optimal performance for a van der Pol limit cycle.

We then move on to the equatorial limit cycle which was originally used to demonstrate phase locking to a semi-classical signal~\cite{Roulet-PRL.121.053601}. 
Despite being insensitive to squeezing, this pure-state limit cycle is shown to outperform the optimally-driven van der Pol oscillator, highlighting the complex interplay between the different quantum resources. 
This understanding leads us to discover a novel type of synchronization blockade based on destructive interference between coherences. 
Finally, we take full advantage of the spin-$1$ Hilbert space and identify the maximum synchronization that can be achieved without imposing any limit cycle nor a specific signal form. 
This fundamental limit is shown to be an asymptotically strict bound that requires (i) a statistical mixture of energy eigenstates in the limit cycle, \emph{i.e.} a larger amplitude uncertainty than that of a pure state, and (ii) a breaking of the symmetry between the extremal spin eigenstates. 
We note that the related question of optimizing the signal to maximize the synchronization of a noisy classical limit cycle is also a subject of research in classical nonlinear dynamics~\cite{Pikovsky-PRL.115.070602}.

This article is structured as follows.
In Sec.\,\ref{sec:Framework} we develop a consistent method to formalize how large the signal strength can be without becoming comparable to the stabilization of the limit cycle. 
This method prepares the ground to quantitatively compare the synchronization behavior of different limit-cycle oscillators. 
Besides discarding the simple tracking of the energy as an indicator of the limit cycle's integrity, our method allows to extend the notion of an Arnold tongue beyond the usual range, revealing a snake-like tongue, which is discussed in Sec.\,\ref{sec:ExtendedArnoldTongue}.
In Secs.\,\ref{sec:VdP} and~\ref{sec:Equatorial} we investigate the spin-$1$ implementations of the van der Pol oscillator and of the equatorial limit cycle, respectively. 
Interference-based quantum synchronization blockade is discussed in Sec.\,\ref{sec:Equatorial:InterferenceBasedSynchronizationBlockade}. 
The bound on maximum synchronization for a spin-$1$ system is derived in Sec.\,\ref{sec:Bound}. 
We discuss the prospects of an experimental observation of quantum synchronization in Sec.\,\ref{sec:Discussion} and conclude in Sec.\,\ref{sec:Conclusions}.

\section{Framework}\label{sec:Framework}
A limit-cycle oscillator is an open system, characterized by a free Hamiltonian $\hat{H}_\text{sys}$, that undergoes a stable periodic motion represented by a closed curve in phase space. 
The stability of this natural rhythm is ensured by the presence of amplitude-dependent gain and damping via a dissipative coupling to an environment. 
In contrast to a coherent drive, such a source of energy does not imprint any preferred phase on the oscillation, thereby allowing the phase of the periodic motion to be freely adjusted by an external perturbation -- the signal -- without affecting the amplitude. This phenomenon is called synchronization.

In this article, we consider the synchronization of a limit-cycle oscillator to an arbitrary external signal of strength $\varepsilon$ that is described by a Hamiltonian $\hat{H}_\text{ext}$. 
This scenario is described by the quantum master equation
\begin{align}
	\dot{\hat{\rho}} = \mathcal{L}_0\hat{\rho} - i \varepsilon\komm{\hat{H}_\text{ext}}{\hat{\rho}} \comma
	\label{eq:masterEq}
\end{align}
where $\hat{\rho}$ is the density matrix of the system and we set $\hbar = 1$. 
This generic equation is the starting point for any study on the synchronization of a single limit-cycle oscillator in the quantum regime.
Actually, it also describes the synchronization of multiple oscillators under a mean-field approximation~\cite{Ludwig-PRL.111.073603,Lee-PRL.111.234101}.
It is typically simulated numerically for a specific limit cycle $\mathcal{L}_0$ and a specific form of the signal $\hat{H}_\text{ext}$, \emph{e.g.} a van der Pol limit cycle subject to a squeezing signal~\cite{Sonar-PRL.120.163601}. 
We will however leave these unspecified for now and instead derive some general properties of the quantum master equation for a limit-cycle oscillator, focusing for simplicity on a spin-$1$ system. 
However, we stress that the methods we introduce in the rest of this section are not tied to this particular platform, but can be readily applied to limit-cycle oscillators living in a different phase space, \emph{e.g.} oscillator-based systems.

\subsection{Spin phase space}
\begin{figure}
	\centering
	\includegraphics[width=0.48\textwidth]{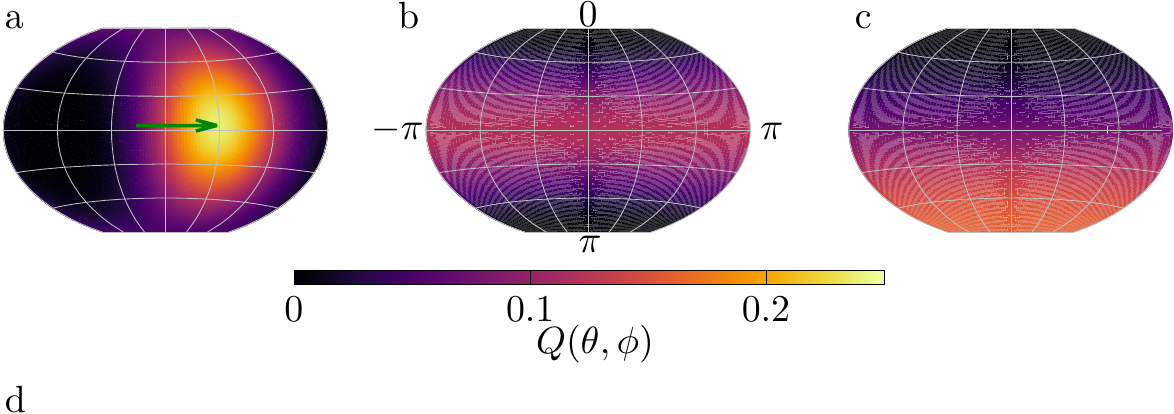}
	\includegraphics[width=0.48\textwidth]{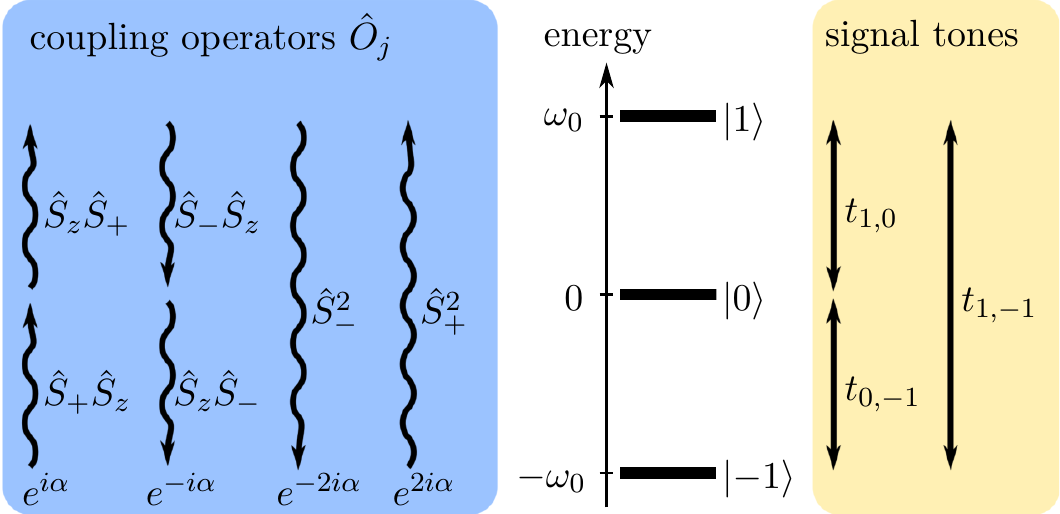}
	\caption{
		(a) -- (c): Illustration of the spherical phase space of a spin-$1$ system. 
		(a) Spin-coherent state $\ket{\theta, \phi} = \ket{\ValueFigureItheta,\ValueFigureIphi}$.
		The green arrow indicates its direction of oscillation in phase space. 
		(b) Equatorial limit cycle $\ket{0}$ considered in Secs.\,\ref{sec:ExtendedArnoldTongue} and~\ref{sec:Equatorial}.
		(c) Limit cycle of the van der Pol oscillator deep in the quantum regime, considered in Sec.\,\ref{sec:VdP}. 
		(d) Energy-level structure of a spin-$1$ system (center), signal tones and corresponding coefficients $t_{i,j}$ of the signal Hamiltonian $\hat{H}_\mathrm{ext}$ (right), and dissipative coupling operators $\hat{O}_j$ that describe unidirectional transitions between two levels (left). 
		The coupling operators are invariant under rotations $\hat{R}_z(\alpha) = e^{-i \alpha \hat{S}_z}$ up to a phase factor indicated in the bottom row. Any linear combination of operators within the same column yields again a valid dissipative coupling operator.
	}
	\label{fig:unit}
\end{figure}

As introduced in Ref.\,\cite{Roulet-PRL.121.053601}, we employ the Husimi function $Q(\theta, \phi \vert \hat{\rho}) = \bra{\theta, \phi} \hat{\rho} \ket{\theta, \phi} 3/4 \pi $ as a phase portrait for spin systems. 
This spherical representation is formulated in terms of spin-coherent states~\cite{Radcliffe-JPhysA.4.313,Arecchi-PhysRevA.6.2211}, which are precisely the states that precess over time according to $\ket{\theta, \phi}\to\ket{\theta, \phi+\omega_0 t}$, as illustrated in Fig.\,\ref{fig:unit}(a), where the natural frequency $\omega_0$ is set by the free Hamiltonian $\hat{H}_\text{sys} = \omega_0 \hat{S}_z$. 
Here $\hat{S}_z$ is the spin component along the quantization axis. 
The azimuth angle $\phi$ thus plays the role of the phase variable at the core of the synchronization formalism, parametrizing the oscillation in phase space.

From the phase-space representation, we can derive the phase distribution $P(\phi \vert \hat{\rho})$ of a given state $\hat{\rho}$ by integrating out the $\theta$ angle, which is analogous to integrating out the amplitude in a position-momentum phase space. 
Since the dissipative source of energy does not favor any phase $\phi$ of the oscillator, the intrinsic quantum noise inevitably leads to phase diffusion such that the limit-cycle state $\hat{\rho}_0$ has a uniform phase distribution $P(\phi \vert \hat{\rho}_0) = 1/2 \pi$, similar to a noisy classical limit-cycle oscillator. 
Therefore, to monitor the phase locking of the limit-cycle oscillator to an external signal we define the shifted phase distribution
\begin{align}
	S(\phi \vert \hat{\rho}) = \int_0^\pi\!\d\theta\, \sin(\theta) Q(\theta, \phi \vert \hat{\rho})  - \frac{1}{2 \pi}\comma
	\label{eqn:Methods:ShiftedPhaseDistribution}
\end{align}
which is identically zero if and only if the distribution is uniform, that is if no phase preference is developed.

\subsection{Limit cycle}
\label{sec:Methods:LC}
Equipped with the phase-space representation reviewed in the previous section, we can now go back to the quantum master equation and derive the form that any limit cycle has to fulfill in a spin-$1$ system, specifying both the available target states as well as the possible responses to perturbations.

The limit cycle is described by the first term in Eq.\,\eqref{eq:masterEq}, which corresponds to the situation without any signal applied, $\varepsilon=0$,
\begin{align}
	\mathcal{L}_0\hat{\rho} = - i \komm{\hat{H}_\text{sys}}{\hat{\rho}} + \sum_{j=1}^N \gamma_j \mathcal{D}[\hat{O}_j] \hat{\rho} \comma 
\end{align}
This dynamics is composed of the oscillation generated by the free Hamiltonian $\hat{H}_\mathrm{sys}$, and of a set of $N$ Lindblad dissipators $\mathcal{D}[\hat{O}] \hat{\rho} = \hat{O} \hat{\rho} \hat{O}^\dagger - \frac{1}{2} \akomm{\hat{O}^\dagger \hat{O}}{\hat{\rho}}$ representing the gain and damping induced by the environment. 
Different choices of coupling operators $\hat{O}_j$ and their corresponding rates $\gamma_j$ define where and how the limit cycle is stabilized in phase space, with the steady state of the dissipative map $\mathcal{L}_0$ being the target state. 
At this point, the fact that the unit can be stabilized in infinitely many ways seems to seriously hinder any attempt to proceed further without focusing on a particular limit cycle. However, we now show that the properties of a limit cycle impose strong constraints on the coupling to the environment, which allows us to narrow down the class of allowed operators and leads to a common structure for valid target states.

The defining feature of a limit cycle is the ability to stabilize the amplitude of the oscillation while leaving the phase completely free. 
The latter is then linearly increasing in time at the natural frequency $\omega_0$ and can be readily adjusted by a weak external signal $\hat{H}_\mathrm{ext}$, possibly to a different frequency. 
We postpone to Sec.\,\ref{sec:Methods:CompareDifferentSystems} the open question of how strong the signal can be without affecting the amplitude of oscillation, and focus here instead on the necessary requirement for the phase to be free before applying a signal.

Specifically, the absence of any phase preference implies that the limit-cycle dynamics generated by $\mathcal{L}_0$ must be invariant under rotations $\hat{R}_z(\alpha) = e^{-i \alpha \hat{S}_z}$ about the axis defined by the free Hamiltonian $\hat{H}_\mathrm{sys}$. 
This is achieved by requiring that the coupling operators $\hat{O}_j$ are themselves invariant up to a phase factor, which does not play any role because of the incoherent nature of the coupling to the environment. 
Hence, the set of allowed operators, shown in Fig.\,\ref{fig:unit}(d), is restricted to those that satisfy $\bra{m} \hat{O}_j \ket{n} \neq 0$ only for a fixed difference $m-n$, where $\ket{n}$ denotes an eigenstate of $\hat{S}_z$. 
Physically, the operators $\hat{O}_j$ correspond to incoherent population transfers that can be combined to stabilize the target state of choice without imposing any phase during the relaxation~\cite{footnote1}. 
An important consequence of the form of the coupling operators is that the dynamics of the limit cycle leads to decoherence in the energy eigenbasis, yielding a diagonal target state. 
In the following, this feature will be key to understand the resource on which quantum synchronization relies.

\subsection{Signal}
Now that we have identified the general form of a limit cycle, the remaining ingredient of Eq.\,\eqref{eq:masterEq} is the external signal which is applied to synchronize the oscillator. 
In a spin-$1$ system, there are up to three transitions that can be externally driven. The corresponding Hamiltonian, in a frame rotating at the signal frequency $\omega_\mathrm{ext}$ and under the rotating-wave approximation, reads
\begin{align} \label{eqn:Methods:Drive}
	\hat{H}_\mathrm{ext} &= t_{0,1} \hat{S}_z \hat{S}_+ - t_{-1,0} \hat{S}_+ \hat{S}_z + t_{-1,1} \hat{S}_+^2 + \mathrm{H.c.}
\end{align}
As illustrated in Fig.\,\ref{fig:unit}(d), it consists of two individual tones applied to the transitions $\ket{-1} \leftrightarrow \ket{0}$ and $\ket{0} \leftrightarrow \ket{1}$, and a squeezing harmonic addressing directly the transition $\ket{1} \leftrightarrow \ket{-1}$. 
The complex parameters $t_{n,m}$ describe the relative phases and amplitudes of these tones. For instance, a semi-classical signal of the form $2 \varepsilon \left[ \cos(\varphi) \hat{S}_x + \sin (\varphi) \hat{S}_y \right]$ corresponds to the first two transitions being equally driven, $t_{0,1} = t_{-1,0} = e^{i \varphi}/2$, and no squeezing tone, $t_{-1,1} = 0$.

\subsection{Perturbation theory}
\label{sec:Methods:PerturbationTheory}
Having fully characterized the spin-$1$ system in terms of the available limit cycles and signals, we now connect the two and develop a concise analytical description of quantum synchronization. 
By definition, synchronization can only be achieved for signal strengths $\varepsilon$ small enough such that the original limit cycle is only weakly perturbed~\cite{Pikovsky-Synchronization}. 
Going beyond this regime would mean affecting not only the phase of the oscillation but its amplitude as well, and thus deforming the limit cycle.
In the following, we refer to this undesired regime as forcing. 
Consequently, we perform an expansion of the density matrix in terms of the signal strength $\hat{\rho}=\sum_{k = 0}^\infty \varepsilon^k \hat{\rho}^{(k)}$, where the first-order term $\hat{\rho}^{(1)}$ contains all the features of synchronization. 
The quantum master equation~\eqref{eq:masterEq} then turns into a set of recursive differential equations~\cite{Li-SciRep.4.4887},
\begin{align}
	\dot{\hat{\rho}}^{(k)} &= \mathcal{L}_0 \hat{\rho}^{(k)} + (1 - \delta_{k,0}) \mathcal{L}_\text{ext} \hat{\rho}^{(k-1)} \comma 
	\label{eqn:PT:QMEExpansion}
\end{align}
with $\mathcal{L}_\text{ext}\hat{\rho} = - i \komm{\hat{H}_\text{ext}}{\hat{\rho}}$ and the normalization condition $\Tr \left[ \hat{\rho}^{(k)} \right] = \delta_{k,0}$.

The leading order $k=0$ corresponds to the situation without any signal being applied. 
As discussed in Section~\ref{sec:Methods:LC}, the system then relaxes to the diagonal steady-state
\begin{align}
	\hat{\rho}^{(0)} = \begin{pmatrix}
		* & 0 & 0 \\
		0 & * & 0 \\
		0 & 0 & *
	\end{pmatrix} \comma \quad \Tr \left[ \hat{\rho}^{(0)} \right] = 1 \comma
	\label{eqn:PT:FormRho0}
\end{align}
where the stars represent non-negative entries that depend on the specific choice of the limit cycle.

The next order $k=1$ accounts for the fact that a weak signal is applied to synchronize the limit-cycle oscillator, yielding the correction $\hat{\rho}^{(1)}$. 
To characterize this term further, we note that the signal Hamiltonian, given in Eq.\,\eqref{eqn:Methods:Drive}, is entirely off-diagonal, $\bra{m} \hat{H}_\mathrm{ext} \ket{n} \propto (1 - \delta_{m,n})$. 
To first order, the signal is thus aiming to generate coherences in the energy eigenbasis. 
On the other hand, we showed that the action of the limit cycle is to equilibrate populations back to the target state $\hat{\rho}^{(0)}$ and, while doing so, to decohere the state in the same basis. 
In matrix form, this means that $\mathcal{L}^0$ takes a block-diagonal structure such that the dynamics of the populations $\bra{n} \hat{\rho} \ket{n}$ and of the coherences $\bra{n} \hat{\rho} \ket{m \neq n}$ are decoupled. 
The block $\mathcal{L}_0^\mathrm{diag}$ acting on the populations is negative-semidefinite, with the vanishing eigenvalue being associated to $\hat{\rho}^{(0)}$, while the block $\mathcal{L}_0^\mathrm{offdiag}$ acting on the coherences has complex eigenvalues with negative real parts that lead to a decay of the coherences.

Going back to the quantum master equation~\eqref{eqn:PT:QMEExpansion}, we thus find that the first order correction
\begin{align}
	\hat{\rho}^{(1)}
	&= - \left( \mathcal{L}_0^{ \mathrm{offdiag}} \right)^{-1} \mathcal{L}_\text{ext} \hat{\rho}^{(0)} \comma \label{eqn:PT:Coherences} 
\end{align}
which is given by the tradeoff between the signal that aims to build up coherences and the limit-cycle dynamics that suppresses them, is purely off-diagonal
\begin{align}
	\hat{\rho}^{(1)} = \begin{pmatrix}
		0 & * & * \\
		* & 0 & * \\
		* & * & 0
	\end{pmatrix} \comma
	\label{eqn:PT:FormRho1}
\end{align}
where the stars represent complex entries compatible with the condition $\hat{\rho}^{(1)\dagger} = \hat{\rho}^{(1)}$. 
This analytical result demonstrates that quantum synchronization achieves phase localization by building up coherences and leaving populations untouched.
The latter is equivalent to preserving the closed curve of the limit cycle in phase space. 
As the signal strength $\varepsilon$ is increased, higher-order corrections contribute where all matrix elements are nonzero in general
\begin{align}
	\hat{\rho}^{(k \geq 2)} = \begin{pmatrix}
		* & * &* \\
		* & * &* \\
		* & * &* 
	\end{pmatrix} \comma \quad \Tr \left[ \hat{\rho}^{(k \geq 2)} \right] = 0 \fullstop
	\label{eqn:PT:FormRho2p}
\end{align}
The coherences driven to first order are now acting back on the populations via the signal Hamiltonian $\hat{H}_\text{ext}$, \emph{e.g.} moving the limit cycle away from its original position in phase space. 
This corresponds to the oscillator being forced.
In the rest of the article, we will restrict the study to the synchronization regime, where higher-order corrections can be neglected,
\begin{align}
	\hat{\rho}\approx \hat{\rho}^{(0)} + \varepsilon \hat{\rho}^{(1)} \fullstop
	\label{eqn:PT:SynchronizationExpansion}
\end{align}

To derive the exact relation between the different coherences that can be built up in the spin-$1$ unit and the resulting localization of the phase, we turn to the phase distribution~\eqref{eqn:Methods:ShiftedPhaseDistribution}, which can be expressed explicitly in terms of the density matrix 
\begin{align}
	S(\phi \vert \hat{\rho}) 
		&= \frac{3}{8 \sqrt{2}} \abs{\rho_{1,0} + \rho_{0,-1}} \cos[\phi + \arg(\rho_{1,0} + \rho_{0,-1})]  \nonumber \\
		&\hphantom{=} + \frac{1}{2 \pi} \abs{\rho_{1,-1}} \cos[2 \phi + \arg(\rho_{1,-1})] \comma 
	\label{eqn:Methods:SynchronizationMeasure}
\end{align}
where $\rho_{n,m} = \bra{n} \hat{\rho} \ket{m}$ are the matrix elements of the state $\hat{\rho}$. 
This is one of the main results of this article. 
A similar formula containing only the $\cos(\phi)$ term has been derived in the specific case of an anharmonic oscillator \cite{Loerch-PhysRevLett.117.073601}.
We first note that $S(\phi \vert \hat{\rho})$ depends only on coherences, and thus on the first order correction $\hat{\rho}^{(1)}$. 
Additionally, the term proportional to $\cos(\phi)$ shows that building up coherences is not a sufficient condition to break the rotational invariance of the limit-cycle state. 
In particular, interference effects between the coherences $\rho_{1,0}$ and $\rho_{0,-1}$ are expected to either enhance or hinder the synchronization behavior. 
We address the latter point in Sec.\,\ref{sec:Equatorial:InterferenceBasedSynchronizationBlockade} where we discuss the possibility of synchronization blockade, despite the energy levels of the spin-$1$ system being equally spaced~\cite{Loerch-PhysRevLett.118.243602}.

Combining Eqs.~\eqref{eqn:PT:SynchronizationExpansion} and~\eqref{eqn:Methods:SynchronizationMeasure} we find that the phase localization increases with the signal strength, $S(\phi \vert \hat{\rho}) = \varepsilon S(\phi \vert \hat{\rho}^{(1)})$. 
On the other hand, we have shown that $\varepsilon$ cannot be increased indefinitely as the system will eventually leave the perturbative regime of synchronization. When comparing the ability of different limit cycles to synchronize to different signals, we thus need a general prescription to set the value of $\varepsilon$ while ensuring that the signal remains a perturbation. 
In the spirit of all past studies which fixed both the signal and the limit cycle, a natural guess would be that normalizing every expression with respect to $\varepsilon$ is sufficient to compare different situations. 
However, since we have kept the signal Hamiltonian \eqref{eqn:Methods:Drive} arbitrary, there are three additional parameters $t_{i,j}$ which determine the relative strength of the signal on each individual transition, as shown in Fig.\,\ref{fig:unit}(d). 
Moreover, each limit cycle has a different response to a given signal, some being deformed earlier than others. 
We are thus required to derive the dimensionless parameter $\eta$ that determines the validity of the first-order approximation~\eqref{eqn:PT:SynchronizationExpansion} in complete generality, which is the subject of the next section.

\subsection{How strong can the signal be?}
\label{sec:Methods:CompareDifferentSystems}
By direct analogy with a classical system, one way to quantify the deformation of a limit cycle is to monitor its change in energy. 
If the signal becomes more than a perturbation, one expects energy to be pumped into the system such that the amplitude of the oscillation is modified and the limit cycle is shifted in phase space. Following this reasoning, the small parameter $\eta$ would then be proportional to the change in the average occupation of the energy levels, which reads for the spin unit
\begin{align}
	p_\text{avg}(\varepsilon) = \Tr \left[\hat{S}_z \left(\hat{\rho}(\varepsilon) - \hat{\rho}^{(0)}\right)\right] \fullstop
\end{align}
As a first sanity check, this deformation measure indeed vanishes in the perturbative regime, where it amounts to evaluate the average occupation of the purely off-diagonal correction $\hat{\rho}^{(1)}$ given in Eq.\,\eqref{eqn:PT:FormRho1}. 
To test it further and check whether it properly detects all types of deformations that can be induced by the signal, we consider a subclass of limit cycles which relax the system to the equatorial state $\hat{\rho}^{(0)} = \ket{0} \bra{0}$. 
The stabilization can be obtained by the two jump operators $\hat{O}_\mathrm{g} = \hat{S}_+ \hat{S}_z$ and $\hat{O}_\mathrm{d} = \hat{S}_- \hat{S}_z$, where the ratio of the associated rates $\gamma_\mathrm{g}/\gamma_\mathrm{d}$ can be freely adjusted to modify the response of the limit cycle to perturbations. 
It is sufficient to restrict ourselves to a semi-classical signal for the rest of this section, \emph{i.e.} $t_{0,1} = t_{-1,0}$ and $t_{-1,1} = 0$.

\begin{figure*}
	\centering
	\includegraphics[width=\textwidth]{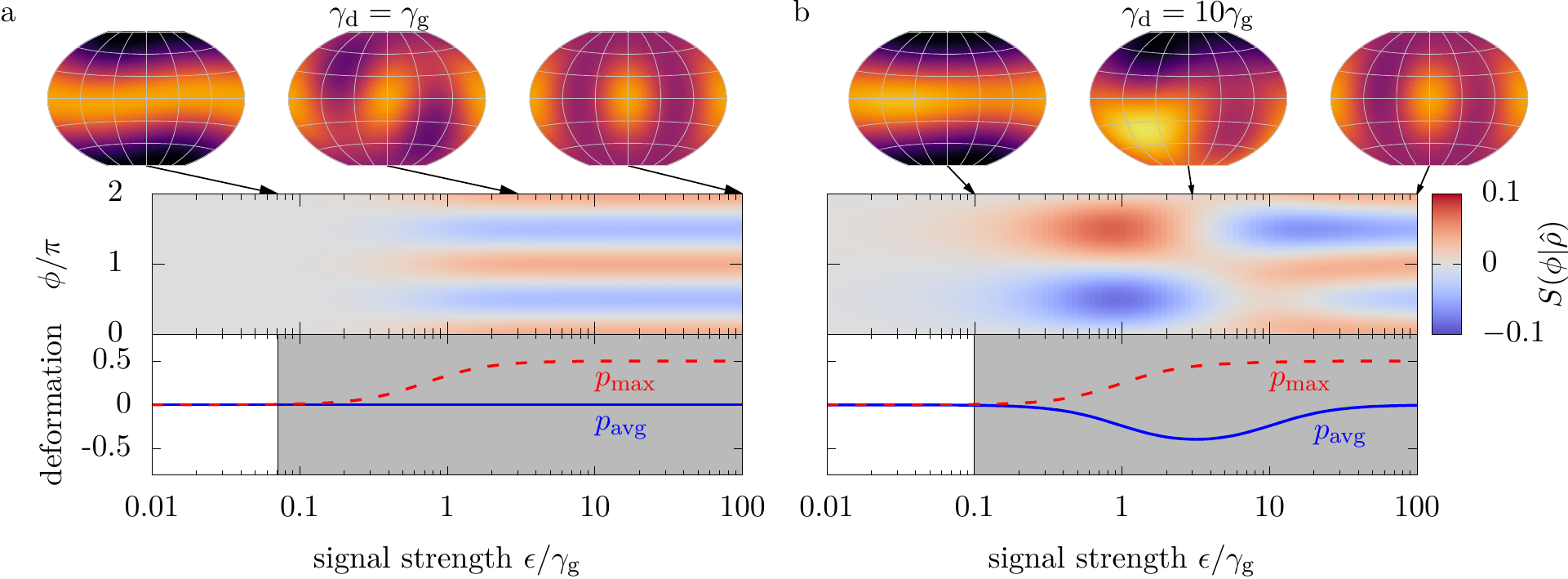}
	\caption{
	Shifted phase distribution $S(\phi \vert \hat{\rho})$ and deformation measures $p_\mathrm{avg}(\varepsilon)$ and $p_\mathrm{max}(\varepsilon)$ as a function of the signal strength $\varepsilon$ for the equatorial limit cycle discussed in Sec.\,\ref{sec:Methods:CompareDifferentSystems} with (a) balanced rates $\gamma_\mathrm{d} / \gamma_\mathrm{g} = \ValueFigureIIgammag$ and (b) imbalanced rates $\gamma_\mathrm{d} /\gamma_\mathrm{g} = \ValueFigureIIgammadimb$. 
	In both cases, a resonant semi-classical signal is applied, \emph{i.e.} $t_{0,1} = t_{-1,0}$, $t_{-1,1} = 0$, and $\Delta = \ValueFigureIIDelta$.
	The gray background in the lower plots indicates the regime of forcing according to Eq.\,\eqref{eqn:Methods:DefinitionEpsilon}, $\varepsilon(\eta) = \eta \gamma_\mathrm{g} \gamma_\mathrm{d}/\sqrt{\gamma_\mathrm{g}^2 + \gamma_\mathrm{d}^2}$, evaluated for $\eta = \ValueFigureIIeta$.
	The plots of the $Q$-function show the state of the system for different values of the signal strength.
	}
	\label{fig:forcing}
\end{figure*}

First focusing on the balanced case $\gamma_\mathrm{g}=\gamma_\mathrm{d}$, Fig.\,\ref{fig:forcing}(a) shows that the signal attracts the phase $\phi$ towards $0$ and $\pi$ without leaving the equator, which seems to be a synchronized state. 
To confirm this visual impression, we track the deformation measure $p_\text{avg}(\varepsilon)$, which stays at zero for the considered range of signal strengths. 
It thus seems that the phase localization is indeed achieved by synchronizing the oscillator to the applied signal. 
Yet, two intriguing features do not agree with this interpretation. Firstly, we have derived in Eq.\,\eqref{eqn:Methods:SynchronizationMeasure} that a synchronized distribution with two stable phases can only emerge by building up coherence between the extremal states, \emph{i.e.} $\rho_{-1,1}\neq 0$, which in turn requires some initial population in the states $\ket{\pm 1}$. 
This is however not possible for the present limit cycle, where only the equatorial state is populated. 
Thus, any synchronized distribution of this limit cycle is predicted to have only a single peak. 
Additionally, if one were to extend the plot range to larger signal strength, the deformation measure would actually be found to vanish for any value of $\varepsilon$. 
This triggers the suspicion that the measure $p_\text{avg}(\varepsilon)$ may not play its role of signaling the transition from the perturbative to the forcing regime for the limit cycle under consideration.

To address this issue, we consider a more fine-grained measure
\begin{align}
	p_\mathrm{max}(\varepsilon) = \max_{n \in \{ -1, 0, 1\}} \abs{\rho_{n,n}(\varepsilon) - \rho_{n,n}^{(0)}} \comma
	\label{eqn:Methods:DeformationMeasurePMax}
\end{align}
which tracks the maximum change of each individual population instead of the averaged $p_\text{avg}(\varepsilon)$. 
As shown in Fig.\,\ref{fig:forcing}(a), this measure is able to detect that the emergence of the two peaks in the phase distribution belongs to the forcing regime. 
Indeed, the onset of the peaks is found to be accompanied by a transfer of population from the equatorial state to the extremal states, which can only be achieved by higher-order contributions~\eqref{eqn:PT:FormRho2p}. 
Due to the symmetry of both the limit cycle $\gamma_\mathrm{d}=\gamma_\mathrm{g}$ and the semi-classical signal $t_{0,1} = t_{-1,0}$, this transfer is however evenly distributed between the extremal states, which explains why the average occupation $p_\text{avg}(\varepsilon)$ remained blind to this deformation.

The balanced limit cycle is thus unable to synchronize to a semi-classical signal. 
Physically, this follows from the fact that to first order the coherences $\rho_{1,0}$ and $\rho_{0,-1}$ are generated with equal amplitudes but opposite sign, and therefore counteract each other in attempting to localize the phase distribution~\eqref{eqn:Methods:SynchronizationMeasure}. 
On the other hand, in the unbalanced case where one of the rates dominates, one of the coherences is able to take the lead and a single-peak phase distribution emerges as illustrated in Fig.\,\ref{fig:forcing}(b). 
This is in agreement with the synchronization reported in Ref.\,\cite{Roulet-PRL.121.053601}. 
Moreover, when the signal is further increased the limit cycle is now clearly deformed towards one of the poles as it enters the forcing regime, before coming back to the equator and forming the same double-peak distribution as in the balanced case.

The results above demonstrate the difficulty of measuring the deformation of a quantum limit cycle based on variations of the populations. 
In fact, there remain some combinations of limit cycle and signal for which even the refined measure $p_\mathrm{max}(\varepsilon)$ is unable to identify the transition to the forcing regime (see Appendix). 
The physical reason for that is that the energy in the finite-dimensional Hilbert space of a spin system is bounded, \emph{i.e.} the amplitude cannot simply grow indefinitely in phase space as the signal strength is increased. 
Hence, there are situations for which the redistribution of the populations in the forcing regime becomes very hard to distinguish from the initial limit-cycle state.

To circumvent this problem, we propose to avoid any coarse-grained deformation measure and instead derive the dimensionless parameter $\eta$ explicitly by requiring that the first-order correction in Eq.~\eqref{eqn:PT:SynchronizationExpansion} remains small with respect to the leading order term, $ \norm{\varepsilon \hat{\rho}^{(1)}} \ll \norm{\hat{\rho}^{(0)}}$. 
Here $\norm{\hat{O}} = \sqrt{\Tr[\hat{O}^\dagger \hat{O}]}$ stands for the Hilbert-Schmidt norm in the operator space, also known as the Liouville space~\cite{fano57}. 
In practice, we impose a fixed threshold value $0 \leq \eta \ll 1$ and set
\begin{align}
	\varepsilon =  \eta \frac{\norm{\hat{\rho}^{(0)}}}{\norm{\hat{\rho}^{(1)}}} \fullstop
	\label{eqn:Methods:DefinitionEpsilon}
\end{align}
The parameter $\eta$ is precisely the expansion parameter that needs to be small to ensure the validity of Eq.\,\eqref{eqn:PT:SynchronizationExpansion}. 
It is also the key ingredient that allows us to compare all sorts of signals and limit cycles, and we end this section by discussing the physical interpretation of Eq.\,\eqref{eqn:Methods:DefinitionEpsilon}.

The numerator, 
\begin{align}
	\norm{\hat{\rho}^{(0)}} = \sqrt{\sum_{m=-1}^1 \abs{\rho^{(0)}_{m,m}}^2} \comma
\end{align}
is similar to the inverse participation ratio used to characterize Anderson localization~\cite{wegner80,evers00}, or to the effective dimension that determines the equilibration of a closed quantum system undergoing unitary dynamics~\cite{linden09,gogolin16}. 
In the three-dimensional Hilbert space of a spin 1, the norm $\norm{\hat{\rho}^{(0)}}$ takes values between $\sqrt{1/3}$, for a limit cycle that is a uniform incoherent mixture of all states, and $1$, for a limit cycle that consists of a single state. 
It captures the fact that a limit cycle with a wider spread of amplitude in phase space is more susceptible to deformations than a narrow limit cycle formed by a single pure state. 
The denominator, on the other hand, is most easily interpreted by assuming that $\mathcal{L}_0^{ \mathrm{offdiag}}$ is diagonalizable and that its eigenoperators $\hat{\mu}_l$, with eigenvalues $\Gamma_l$, form an orthonormal basis spanning the space of coherences. 
Expressing the impact of the signal in this basis with the projection coefficients $g_l=\Tr[\hat{\mu}_l^\dagger\mathcal{L}_\text{ext} \hat{\rho}^{(0)}]$, we can then rewrite the first-order term \eqref{eqn:PT:Coherences} as $\hat{\rho}^{(1)} = - \sum_l \hat{\mu}_l (g_l/\Gamma_l) $ and obtain for the norm
\begin{align}
	\norm{\hat{\rho}^{(1)}} = \sqrt{\sum_l \abs{\frac{g_l}{\Gamma_l}}^2} \fullstop
\end{align}
The decomposition coefficients $g_l$ describe how strongly a certain eigencoherence is driven away from zero by the signal $\mathcal{L}_\mathrm{ext}$, and are compared to the corresponding relaxation rates $\Gamma_l$. 
Hence, the denominator of Eq.\,\eqref{eqn:Methods:DefinitionEpsilon} ensures that the overall effect of the signal on each eigencoherence remains small compared to the stabilization of the limit cycle.

Note that the assumptions that $\mathcal{L}_0^{ \mathrm{offdiag}}$ is diagonalizable and that the eigencoherences form an orthonormal basis have only been used to discuss the physical meaning of the threshold $\eta$. 
In particular, the definition~\eqref{eqn:Methods:DefinitionEpsilon} remains well-defined even if these simplifying assumptions do not hold.

\section{Extended Arnold tongue}
\label{sec:ExtendedArnoldTongue}
For the rest of the article, we consider the maximum of the shifted phase distribution as a single-number measure of synchronization \cite{Lee-PRL.111.234101,Roulet-PRL.121.053601,Loerch-PhysRevLett.118.243602}, 
\begin{align}
	\mathcal{S}(\hat{\rho}) = \max_{\phi \in [0, 2 \pi)} \varepsilon S(\phi \vert \hat{\rho}^{(1)}) \fullstop
	\label{eqn:Methods:SynchronizationMeasureMax}
\end{align}

As a first application of the formalism developed in the previous section, we address the open question of delimiting the synchronization region as a function of the detuning $\Delta=\omega_0-\omega_\mathrm{ext}$ and the signal strength $\varepsilon$. 
It is known that the range of detunings for which synchronization survives increases with the signal strength~\cite{Pikovsky-Synchronization}. 
This yields the classic triangular region called the Arnold tongue, which is typically plotted up to an arbitrary signal strength $\varepsilon_\text{max}(0)$ that is qualitatively chosen to ensure that the signal is only weakly perturbing the limit cycle for any value of the detuning~\cite{Sonar-PRL.120.163601,Roulet-PRL.121.053601,Roulet-PRL.121.063601}.

\begin{figure}
	\centering
	\includegraphics[width=0.48\textwidth]{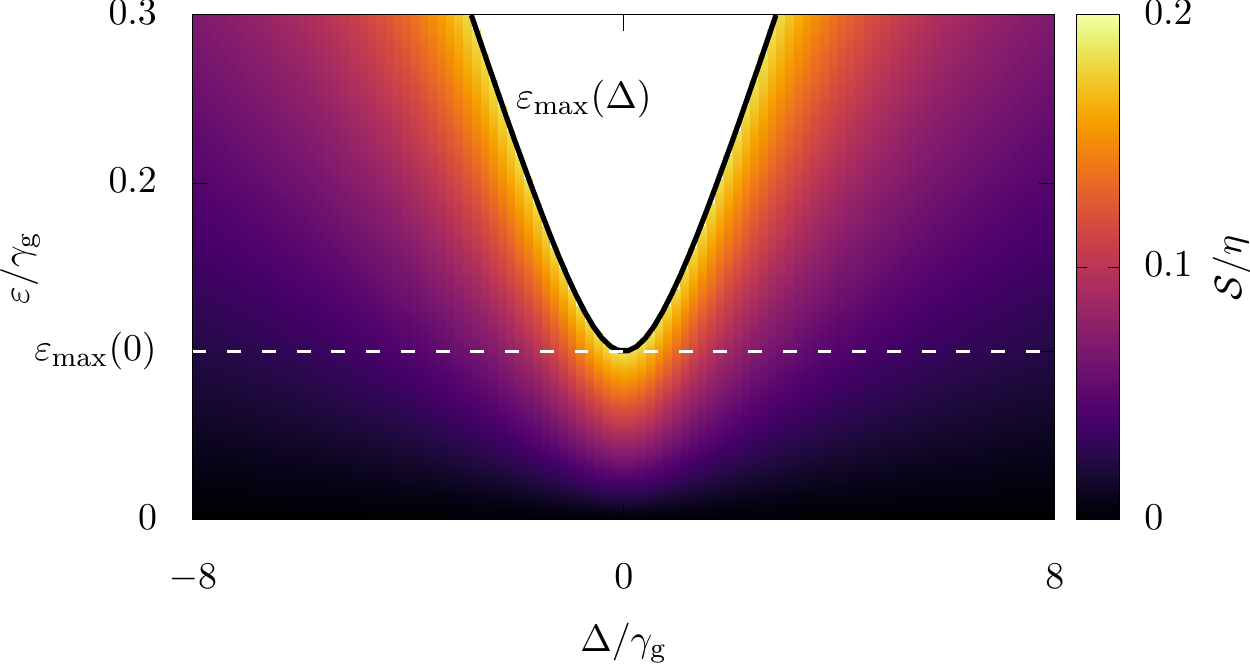}
	\caption{
		Extended Arnold tongue for the equatorial limit-cycle oscillator introduced in Sec.\,\ref{sec:Methods:CompareDifferentSystems} with imbalanced dissipation rates, $\gamma_\mathrm{d}/\gamma_\mathrm{g} = \ValueFigureIIIgammad$, subject to a semi-classical signal, $t_{0,1} = t_{-1,0}$, $t_{-1,1} = 0$. 
		Usually, the Arnold tongue is plotted for any detuning $\Delta$ up to a maximum cutoff value $\varepsilon_\mathrm{max}(0)$, indicated here by a dashed white line.
		Our method allows us to derive the boundary between the synchronization regime (colored) and the forcing regime (white) as a function of the detuning, $\varepsilon(\Delta) = \eta/\sqrt{(\gamma_\mathrm{d}^2 + \Delta^2)^{-1} + (\gamma_\mathrm{g}^2 + \Delta^2)^{-1}}$, which is represented by the solid black line. 
		The Arnold tongue is extended for nonzero detuning and becomes a snake-like split tongue. 
		The threshold is $\eta = \ValueFigureIIIeta$. 
	}
	\label{fig:2:Arnold}
\end{figure}

Our method allows us to proceed further and to formally derive the analytical boundary by explicitly tracking the validity of the perturbation theory for a fixed threshold $\eta$. 
Figure\,\ref{fig:2:Arnold} illustrates this result for the equatorial limit cycle introduced in the previous section: we can indeed obtain the maximum signal that is permitted on resonance $\varepsilon_\text{max}(0)$, which determines the optimal horizontal cut of the tongue. 
However, we find that the boundary of the synchronization region is actually a function of the detuning, thereby demonstrating that the standard horizontal cut is discarding an entire part of the Arnold tongue. 
The physical origin of this uncharted region is that the ability of the signal to affect the unit, \emph{i.e.} to drive coherences, is reduced as the detuning is increased. 
To compensate this loss in susceptibility of the unit, the signal strength can thus be increased beyond the resonant bound, $\varepsilon_\text{max}(\Delta)\geq\varepsilon_\text{max}(0)$. 
To our knowledge, this is the first time that the Arnold tongue is extended to larger off-resonant drive strengths, yielding a snake-like split tongue.

\section{Van der Pol limit cycle}
\label{sec:VdP}
The van der Pol oscillator has been proposed a century ago as a tool to gain theoretical insight into the phenomenon of synchronization~\cite{Pikovsky-Synchronization}. 
After the success of the model in the classical world, it has recently been quantized and studied in the regime of a few excitations to probe numerically the features of quantum synchronization~\cite{Lee-PRL.111.234101,Walter-PRL.112.094102}. 
Coming back to the spin-$1$ system under study, it may not be clear at first sight whether any link can be drawn between a mathematical model formulated within the position-momentum phase space of an oscillator and a purely quantum system with no classical analogue. 
However, we now show that when operated deep in quantum regime, the van der Pol limit cycle can be faithfully represented in the spin-$1$ system, which grants access to tractable analytics and demonstrates the versatility of the most elementary quantum unit to study quantum synchronization.

\subsection{Harmonic oscillator vs. spin 1}
\label{sec:vdP:HOPhaseSpace}
The defining characteristic of the van der Pol model is the stabilization of the self-sustained oscillations, which is achieved by a linear gain acting against a nonlinear damping. 
In the weakly-nonlinear regime where the limit cycle is essentially circular in phase space, the quantum counterpart of this dissipative dynamics is realized for harmonic oscillators, $\hat{H}_\mathrm{sys} = \omega_0 \hat{a}^\dagger \hat{a}$, by a single-photon gain $\hat{O}_\mathrm{g} = \hat{a}^\dagger$ and a two-photon loss $\hat{O}_\mathrm{d} = \hat{a}^2$~\cite{Lee-PRL.111.234101,Walter-PRL.112.094102}. 
Bringing the oscillator in the quantum regime then amounts to increasing the damping so that occupied Fock states are strongly relaxed towards the bottom of the energy ladder, except for the first excited state, which is unaffected by the two-photon loss. 
Accordingly, the oscillator is confined in the vicinity of the first excited state and mostly couples to the vacuum and the two-photon Fock state when submitted to a weak signal.
Hence, deep in the quantum regime, where the van der Pol oscillator is effectively restricted to the three lowest Fock states~\cite{Lee-PRL.111.234101,Walter-PRL.112.094102}, the three levels of our spin-$1$ system provide a valid support.

To implement the dissipative dynamics in the spin platform, we consider the single excitation gain $\hat{O}_\mathrm{g} = \hat{S}_z \hat{S}_+ - \hat{S}_+ \hat{S}_z/\sqrt{2}$ and the two-excitation loss $\hat{O}_\mathrm{d} = \hat{S}_-^2/\sqrt{2}$, with respective rates $\gamma_\mathrm{g}$ and $\gamma_\mathrm{d}$. 
This specific form is chosen such that the matrix representations of $\hat{O}_\mathrm{g}$ and $\hat{O}_\mathrm{d}$ are identical to the matrix representations of the creation $\hat{a}^\dagger$ and two-photon annihilation $\hat{a}^2$ operators of an oscillator restricted to the three lowest Fock states. 
Similarly, we renormalize the signal coefficients for the rest of the section as follows
\begin{align}
   t_{0,1} &= \tau_{0,1} \comma \nonumber\\
   t_{-1,0} &= \tau_{-1,0} / \sqrt{2} \comma \\
   t_{-1,1} &= \tau_{-1,1} / \sqrt{2} \fullstop\nonumber
\end{align}

Having specified the stabilization of the limit cycle, we obtain the steady-state populations by solving the leading-order quantum master equation~\eqref{eqn:PT:QMEExpansion}, \emph{i.e.} $\mathcal{L}_0\hat{\rho}^{(0)}=0$, which yields
\begin{align}
   \hat{\rho}^{(0)}_{1,1} &= \frac{\gamma_\mathrm{g}}{3 \gamma_\mathrm{d} + \gamma_\mathrm{g}} \comma \nonumber\\
   \hat{\rho}^{(0)}_{0,0} &= \frac{\gamma_\mathrm{d}}{3 \gamma_\mathrm{d} + \gamma_\mathrm{g}} \comma \\\nonumber
   \hat{\rho}^{(0)}_{-1,-1} &= \frac{2 \gamma_\mathrm{d}}{3 \gamma_\mathrm{d} + \gamma_\mathrm{g}} \fullstop
\end{align}
In the regime of interest $\gamma_\mathrm{d} \gg \gamma_\mathrm{g}$, the populations converge to the values ($0,1/3,2/3$), which are precisely those of a van der Pol limit cycle implemented in a harmonic oscillator~\cite{Lee-PRL.111.234101,Walter-PRL.112.094102}. 
Hence, as long as the oscillator is indeed confined deep in the quantum regime, its effective density matrix truncated to the first three levels of the harmonic ladder is identical to that of a spin-based van der Pol oscillator. 
Since the perturbation expansion~\eqref{eqn:PT:QMEExpansion} is valid for both systems, the equivalence remains true once a signal $\mathcal{L}_\mathrm{ext}$ is applied. 
Conversely, any difference between the states of the two platforms indicates that the oscillator is transitioning towards the classical regime, populating higher Fock states, and thus losing the possibility to be represented in a spin-$1$ system.

To conclude the comparison, we note that there remains a fundamental difference between the two architectures, namely the phase space representation which is at the core of the synchronization phenomenon. 
Specifically, the infinite position-momentum plane of a harmonic oscillator is replaced by a sphere, that is a space of different topology. 
To derive the impact of this change on the measure of phase localization, we employ the counterpart of the spin phase distribution~\eqref{eqn:Methods:ShiftedPhaseDistribution} for an oscillator~\cite{Gerry-QO,armour15}
\begin{align}
   S_\mathrm{osc}(\phi \vert \hat{\rho}) = \frac{1}{2 \pi} \bra{\phi} \hat{\rho} \ket{\phi} - \frac{1}{2 \pi} \comma
\end{align}
where $\ket{\phi}$ is a phase state defined in terms of the three lowest Fock states $\ket{n_\mathrm{F}}$, $n_\mathrm{F} \in \{0, 1, 2\}$
\begin{align}
   \ket{\phi} = \sum_{n_\mathrm{F}=0}^2 e^{i n_\mathrm{F} \phi} \ket{n_\mathrm{F}} \fullstop
\end{align}
Expressed in terms of the density-matrix elements $\rho_{n_\mathrm{F},m_\mathrm{F}} = \bra{n_\mathrm{F}} \hat{\rho} \ket{m_\mathrm{F}}$, it takes the form
\begin{align}
   S_\mathrm{osc}(\phi \vert \hat{\rho})
   &= \frac{1}{2 \pi} \abs{\rho_{1,0} + \rho_{2,1}} \cos [\phi + \arg( \rho_{1,0} + \rho_{2,1})] \nonumber \\
   &+ \frac{1}{2 \pi} \abs{\rho_{2,0}} \cos[2 \phi + \arg(\rho_{2,0})] \comma
   \label{eqn:vdP:SyncMeasure}
\end{align}
which differs from Eq.\,\eqref{eqn:Methods:SynchronizationMeasure} only in the constant preceding the $\cos(\phi)$ term. 
Therefore, the qualitative synchronization behavior of the van der Pol model is identical in both platforms and we can exploit the spin system to characterize the deep quantum regime analytically. 
For the rest of this section, we assume $\gamma_\mathrm{d} \gg \gamma_\mathrm{g}, \Delta$ unless stated otherwise.

\subsection{Semi-classical and squeezing signal}
\begin{figure}
	\centering
	\includegraphics[width=0.48\textwidth]{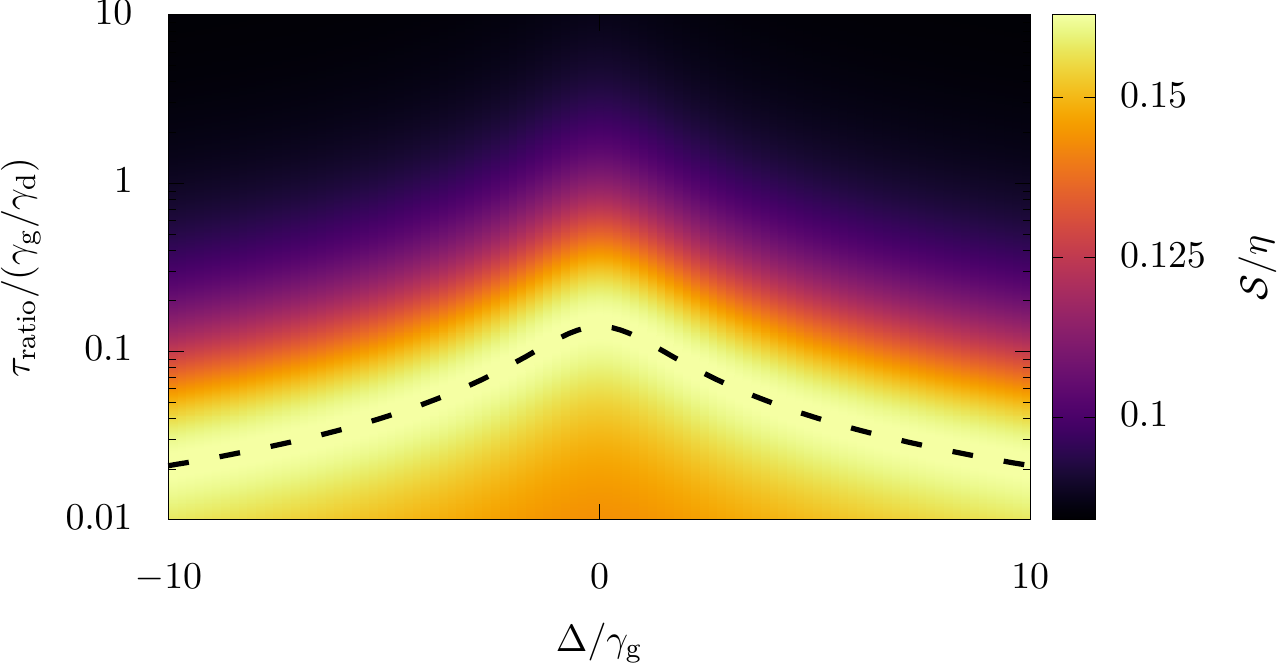}
	\caption{
		Synchronization of the van der Pol oscillator deep in the quantum regime to a combination of semi-classical and squeezing tones at relative strength $\tau_\mathrm{ratio} = \abs{\tau_{-1,1}}/\abs{\tau_{0,1}}$.
		The color bar ranges from the minimum synchronization $\sqrt{5/2}/6 \pi$ achieved for $\tau_\mathrm{ratio} \to \infty$ to the maximum synchronization $\mathcal{S}/\eta = \sqrt{5(32 + 9 \pi^2)}/48 \pi$ at the optimal ratio $\tau_\mathrm{ratio}^\mathrm{opt}$, which is indicated by the dashed black line. 
		Parameters are $\gamma_\mathrm{d}/\gamma_\mathrm{g} = \ValueFigureIVgammad$ and $\eta = \ValueFigureIVeta$. 
	}
	\label{fig:4}
\end{figure}

We start by considering a situation explored in a recent numerical study, which showed that the synchronization of a van der Pol oscillator can be significantly enhanced by exploiting the quantumness of the system, specifically by adding a squeezing tone to a semi-classical signal~\cite{Sonar-PRL.120.163601}. 
In the spin system considered here, this corresponds to fixing the signal tones as $\tau_{0,1} = \tau_{-1,0}$ and $\tau_{-1,1}\neq 0$. 
In addition, we adjust the relative phase between the squeezing tone and the semi-classical component such that they aim at localizing the same phase, \emph{i.e.} such that the $\cos(\phi)$ and $\cos(2\phi)$ terms in Eq.\,\eqref{eqn:Methods:SynchronizationMeasure} share a common peak. 
As shown in Fig.\,\ref{fig:4}, the resulting measure of synchronization $\mathcal{S}/\eta$ corroborates the numerical findings of Ref.\,\cite{Sonar-PRL.120.163601} near resonance, namely, the van der Pol oscillator synchronizes better to signals dominated by a squeezing tone $\tau_\text{ratio}=|\tau_{-1,1}|/|\tau_{0,1}|\gg 1$. 
However it seems that this advantage is substantially reduced, if not suppressed, when trying to lock to an off-resonant signal. 
There, the semi-classical component should be favored in order to maximize the phase localization.

To investigate this tradeoff and establish whether squeezing is only beneficial within a narrow bandwidth around resonance, we turn to analytics and derive the first-order correction $\hat{\rho}^{(1)}$ via Eq.\,\eqref{eqn:PT:Coherences}. 
Substituting the obtained state into the definition of the synchronization measure~\eqref{eqn:Methods:SynchronizationMeasureMax}, we find deep in the quantum regime the compact form
\begin{align}
   \mathcal{S} = \eta\frac{\sqrt{5}}{48\pi}\frac{3\pi\gamma_\mathrm{d}+ 8\tau_\text{ratio}\sqrt{9\gamma_\mathrm{g}^2+4\Delta^2}}{\sqrt{\gamma_\mathrm{d}^2+ 2\tau_\text{ratio}^2 \left(9\gamma_\mathrm{g}^2+4\Delta^2\right)}} \fullstop
\end{align}
Indeed, the maximum synchronization $\mathcal{S}/\eta  =\sqrt{5(32+9\pi^2)}/48\pi\approx 0.163$ is achieved by the optimal squeezing ratio $\tau_\text{ratio}^\text{opt}= 4\gamma_\mathrm{d} (9\gamma_\mathrm{g}^2+4\Delta^2)^{-1/2}/3\pi$ which decreases with the detuning. 
On the other hand, synchronization to a purely semi-classical signal without the squeezing tone, $\tau_\text{ratio}=0$, is limited to $\mathcal{S}/\eta=\sqrt{5}/16\approx 0.140$. 
The access to a squeezing tone on top of a semi-classical one is thus always beneficial for the van der Pol limit cycle.
However, note that synchronization decreases again in the limit $\tau_\mathrm{ratio} \to \infty$, where we find $\mathcal{S}/\eta\to\sqrt{5/2}/6\pi\approx 0.084$.

\subsection{Optimized signal}\label{sec:vdpOpt}
In the previous section, we have reproduced results that had previously been obtained with harmonic oscillators, and we have demonstrated the power of the spin-$1$ platform to go beyond numerics using the formalism developed in this manuscript. 
We now conclude our study of the van der Pol limit cycle by answering the fundamental question of what is the maximum synchronization that can be achieved for a van der Pol oscillator deep in the quantum regime.

To this end, we relax the semi-classical restriction $\tau_{0,1} = \tau_{-1,0}$ and employ the following parametrization
\begin{align}
  	\tau_{0,1} &= c\cos(\zeta) e^{i \chi} \comma 
      \label{eqn:vdP:GeneralSignal:Parametrization} \\
      \tau_{-1,0} &= c \sin(\zeta) \comma \nonumber
\end{align}
with $c>0$, $\tau_\text{ratio}= \abs{\tau_{-1,1}}/c$, $0 \leq \zeta \leq \pi/2$, and $0 \leq \chi \leq 2 \pi$. 
Using Eq.\,\eqref{eqn:PT:Coherences}, we compute the first order correction $\hat{\rho}^{(1)}$ and obtain the synchronization measure $\mathcal{S}(\hat{\rho}^{(1)})$ for any choice of parameters. 
We omit here the general formula, which is rather lengthy and uninformative as such. 
Instead, we perform an exhaustive optimization over all three signal tones, focusing on the resonant case $\Delta=0$ for simplicity.

\begin{figure}
	\centering
	\includegraphics[width=0.48\textwidth]{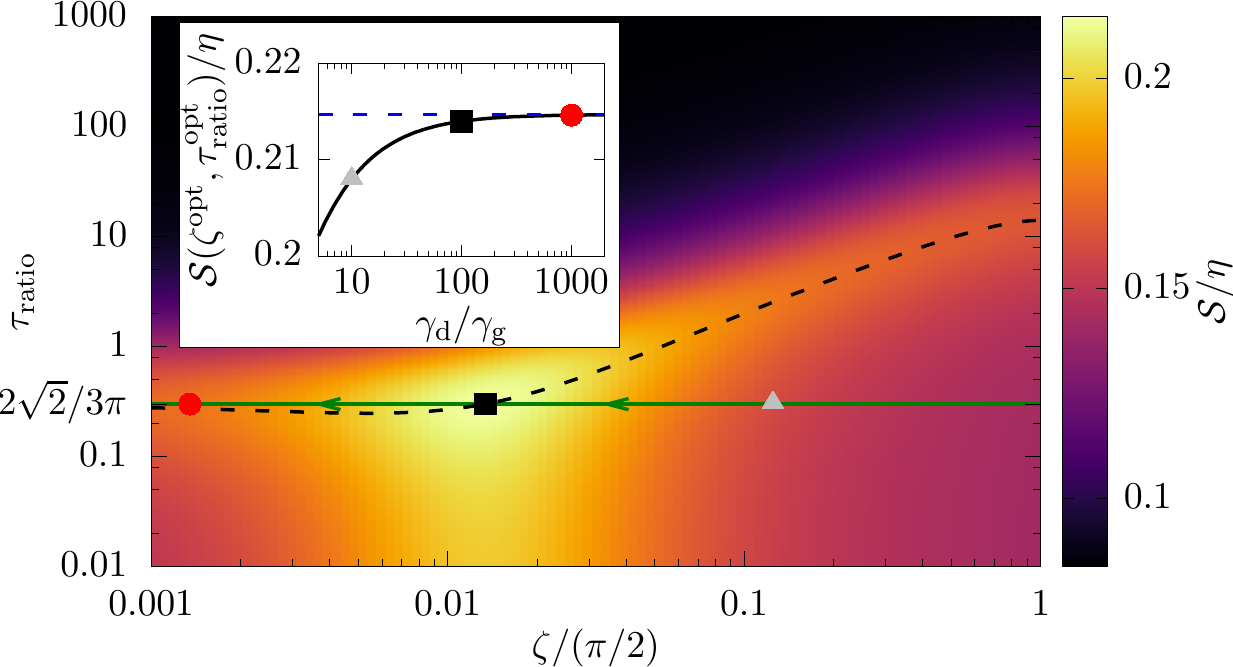}
	\caption{
		Synchronization of the van der Pol oscillator deep in the quantum regime, $\gamma_\mathrm{d}/\gamma_\mathrm{g} = \ValueFigureIVbgammad$, to a general signal~\eqref{eqn:vdP:GeneralSignal:Parametrization} with $\chi = 0$. 
		For reference, the dashed black line indicates the optimal ratio of squeezing for a fixed $\zeta$.
		The solid markers indicate from right to left the optimal signal parameters for $\gamma_\mathrm{d}/\gamma_\mathrm{g} = \ValueFigureIVbgammadC$, $\ValueFigureIVbgammad$, and $\ValueFigureIVbgammadB$.  
		In the quantum regime, the optimal value of $\tau_\mathrm{ratio}$ converges to $\tau_\mathrm{ratio}^\mathrm{opt} = 2 \sqrt{2}/3 \pi$ and $\zeta^\mathrm{opt}$ decreases with $\gamma_\mathrm{d}/\gamma_\mathrm{g}$, as indicated by the solid green arrows. 
		Inset: $\mathcal{S}/\eta$ evaluated at the optimal values $\tau_\mathrm{ratio}^\mathrm{opt}$ and $\zeta^\mathrm{opt}$ as a function of $\gamma_\mathrm{d}/\gamma_\mathrm{g}$. 
		Maximum synchronization is obtained in the limit $\gamma_\mathrm{d}/\gamma_\mathrm{g} \to \infty$ where $\mathcal{S}/\eta$ converges to $\sqrt{40 + 45 \pi^2/2}/24 \pi \approx 0.215$, indicated by the dashed blue line. 
		The threshold parameter is $\eta = \ValueFigureIVbeta$. 
	}
	\label{fig:5}
\end{figure}

In the resonant case, the optimal phase of the semi-classical signal components is $\chi = 0$. 
As illustrated in Fig.\,\ref{fig:5}, we find that maximum synchronization deep in the quantum regime is achieved for $\zeta^\mathrm{opt} = \arccot( \sqrt{2} \gamma_\mathrm{d}/3 \gamma_\mathrm{g})$ and $\tau_\mathrm{ratio}^\mathrm{opt} = 2 \sqrt{2}/3 \pi$.
However, similar to the situation encountered in the previous section, note that the tone $\tau_{-1,0}$ cannot be simply switched off, $\zeta=0$, because in this case the synchronization is limited to $\mathcal{S}/\eta=\sqrt{5(32 + 9\pi^2)}/48\pi \approx 0.163$.
As displayed in the inset of Fig.\,\ref{fig:5},
the maximum synchronization that is possible for a van der Pol limit cycle takes the value 
\begin{align}
   \mathcal{S}/\eta = \frac{\sqrt{40 + \frac{45}{2} \pi^2}}{24 \pi} \approx 0.215 \fullstop
\end{align}
This is one of the main results of the article, which will allow us to compare the van der Pol model with other limit cycles available in the spin-$1$ system.

\section{Equatorial limit cycle}
\label{sec:Equatorial}
Moving away from classically-inspired limit cycles, we consider in this section the equatorial limit cycle used in Sec.\,\ref{sec:Methods:CompareDifferentSystems} and defined by the dissipative coupling operators $\hat{O}_\mathrm{g} = \hat{S}_+ \hat{S}_z$ and $\hat{O}_\mathrm{d} = \hat{S}_- \hat{S}_z$, with respective rates $\gamma_\mathrm{g}$ and $\gamma_\mathrm{d}$. 
The key feature of the resulting stabilization is its simplicity, because the extremal states $\ket{\pm1}$ are independently relaxed to the equatorial state
\begin{align}
	\hat{\rho}^{(0)}=\ket{0}\bra{0} \fullstop
\end{align}
Incidentally, the absence of initial population in the extremal states $\rho^{(0)}_{\pm1,\pm1}=0$ renders the limit cycle insensitive to a squeezing signal, such that $\rho_{-1,1}^{(1)}$ is bound to stay zero. 
However, the remaining coherences, 
\begin{align}
	\rho_{1,0}^{(1)} &= - \frac{i \sqrt{2}}{\gamma_\mathrm{d} + i \Delta} t_{0,1} \comma \nonumber \\
	\rho_{0,-1}^{(1)} &= + \frac{i \sqrt{2}}{\gamma_\mathrm{g} + i \Delta} t_{-1,0} \comma
	\label{eqn:Equatorial:Coherences}
\end{align}
can be built up independently by the signal tones of the corresponding transition. 
Therefore, we can directly exploit their impact on $\mathcal{S}$, as given by Eq.\,\eqref{eqn:Methods:SynchronizationMeasure}, and we find that a straightforward combination of the semi-classical signal tones outperforms the maximal synchronization achieved by a van der Pol limit cycle.

To proceed further, we choose the following parametrization of the signal
\begin{align}
	t_{0,1} &= \cos(\zeta) e^{i \chi} \comma \nonumber\\
	t_{-1,0} &= \sin(\zeta) \comma \label{eqn:Equatorial:SignalParametrization}\\\nonumber
	t_{-1,1} &= 0 \fullstop
\end{align}
Remarkably, this time the synchronization measure can be expressed in a compact form without imposing any constraint on the signal, 
\begin{align}
	\mathcal{S} &= \eta \frac{3}{16} \sqrt{1 - 2 \frac{\sin(\zeta) \cos(\zeta) \cos(\chi + \alpha)}{r \cos^2(\zeta) + \frac{1}{r} \sin^2(\zeta)}} \comma \nonumber\\
	r &= \sqrt{\frac{\gamma_\mathrm{g}^2 + \Delta^2}{\gamma_\mathrm{d}^2 + \Delta^2}} \comma \label{eqn:Equatorial:Smax}\\\nonumber
	\alpha &= \arg \left( \frac{1}{\gamma_\mathrm{g} - i \Delta} \frac{1}{\gamma_\mathrm{d} + i \Delta} \right) \fullstop
\end{align}

\subsection{Semi-classical signal}
First we analyze synchronization to a semi-classical signal, $t_{0,1}=t_{-1,0}$, parametrized by $\chi = 0$ and $\zeta = \pi/4$. 
This corresponds to the scenario studied in Ref.\,\cite{Roulet-PRL.121.053601}, where synchronization was found to vanish for balanced dissipation rates $\gamma_\mathrm{d} = \gamma_\mathrm{g}$. 
Within the present framework, we can go a step further and identify the physical origin of this singularity: for balanced rates, the semi-classical signal is building up both coherences with the same strength against the same relaxation rate, yielding the same absolute value but with opposite phase $\rho_{0,1}= - \rho_{-1,0}$. 
Since the synchronization measure~\eqref{eqn:Methods:SynchronizationMeasure} is a function of their sum, $\abs{\rho_{0,1} + \rho_{-1,0}}$, this leads to destructive interference, captured by the factor $\cos(\chi + \alpha)=1$ in Eq.\,\eqref{eqn:Equatorial:Smax}, and no synchronization is observed.

Building on this understanding, we find that for any finite asymmetry between the rates, one of the coherences dominates such that the impact of the destructive interference is reduced. 
The synchronization is then maximal on resonance $\Delta = 0$, where Eq.\,\eqref{eqn:Equatorial:Smax} takes the form
\begin{align}
	\mathcal{S} = \eta \frac{3}{16} \sqrt{1 - \frac{2\gamma_\mathrm{d} \gamma_\mathrm{g}}{\gamma_\mathrm{d}^2 + \gamma_\mathrm{g}^2}} \fullstop
	\label{eqn:Equatorial:semiclassicalS}
\end{align}
In particular, large asymmetries yield the maximum synchronization for a semi-classical signal $\mathcal{S}/\eta = 3/16\approx 0.188$, where only one of the coherences contributes without being suppressed by the other. 
Comparing with the van der Pol limit cycle, this value is larger than the one obtained for the same signal, $\mathcal{S}/\eta \approx 0.140$, but lower than for the optimized signal, $\mathcal{S}/\eta \approx 0.215$, which exploited all three coherences.

\subsection{Optimized signal}
\label{eq:equaOpt}
\begin{figure}
	\centering
	\includegraphics[width=0.48\textwidth]{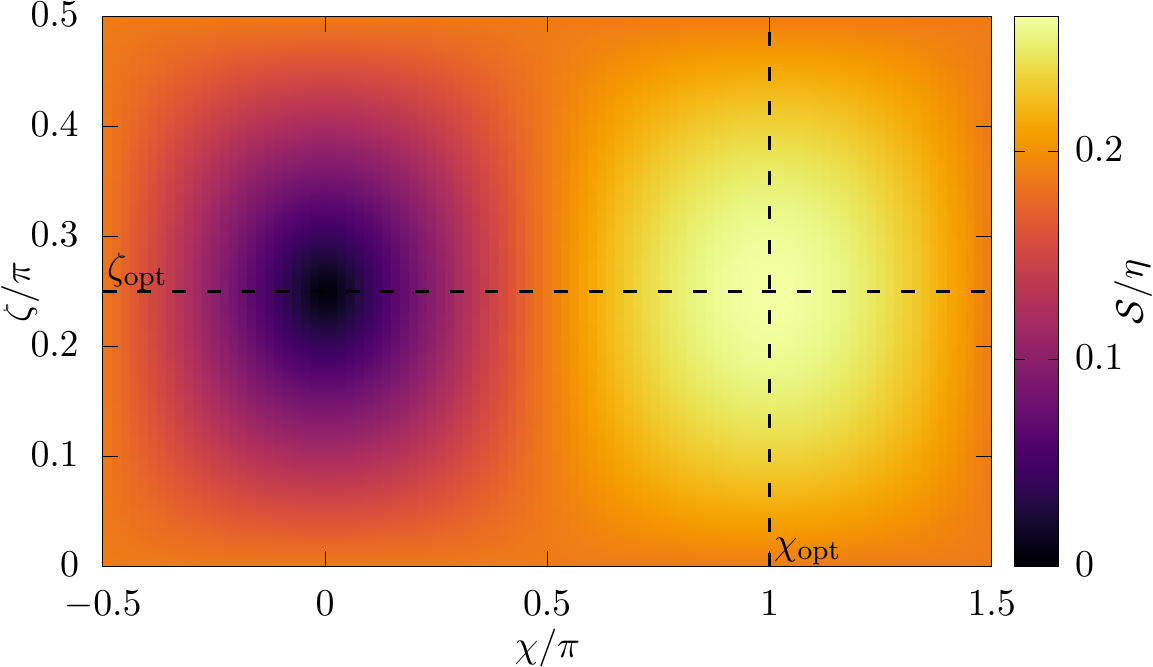}
	\caption{\label{fig:5:Equatorial:2D}
		Equatorial limit cycle with balanced dissipation rates $\gamma_\mathrm{g} = \gamma_\mathrm{d}$ subject to a resonant signal parametrized by the convention of Eq.\,\eqref{eqn:Equatorial:SignalParametrization}. 
		The relative phase $\chi$ between the signal tones determines the relative phase of the coherences $\rho_{0,1}$ and $\rho_{-1,0}$. 
		They interfere constructively for $\chi^\mathrm{opt} = \pi$ and destructively for $\chi = 0$. 
		The parameter $\zeta$ determines if the amplitudes of the two tones are equal ($\zeta^\mathrm{opt} = \pi/4$) or different. 
		The maximum synchronization for the equatorial limit cycle, $\mathcal{S}/\eta = 3 \sqrt{2}/16 \approx 0.265$, is obtained at the intersection of the dashed black lines where both coherences have the same amplitude and interfere constructively. 
		A semi-classical signal corresponds to $\chi = 0$. 
		The threshold parameter is $\eta = \ValueFigureVeta$. 		
	}
\end{figure}

The strength of the equatorial limit cycle is the possibility to address the coherences individually. 
To improve on the van der Pol model, we thus aim for a signal where the coherences are built in phase and therefore interfere constructively. 
At the level of the synchronization measure~\eqref{eqn:Equatorial:Smax}, this amounts to requiring that $\cos(\chi + \alpha)=-1$. 
We are then left with the task of maximizing the term $2 \sin(\zeta) \cos(\zeta)/[\cos^2(\zeta) r + \sin^2(\zeta)/r]$.
This yields the optimal angles
\begin{align}
	\chi^\mathrm{opt} &= \pi- \alpha \comma \nonumber\\
	\zeta^\mathrm{opt} &= \arctan(r) \comma
	\label{eq:optEq}
\end{align}
where the second condition implies that both coherences have the same amplitude. 
The resulting constructive interference yields
\begin{align}
	\frac{\mathcal{S}}{\eta} = \frac{3}{16} \sqrt{2} \approx 0.265\comma
	\label{eq:maxEq}
\end{align}
which is the maximum synchronization that is possible for the equatorial limit cycle and which outperforms the capabilities of the van der Pol limit cycle. 
This result is illustrated in Fig.\,\ref{fig:5:Equatorial:2D} for the case of balanced dissipation rates, where synchronization to a semi-classical signal is not possible.

\section{Interference-based synchronization blockade}
\label{sec:Equatorial:InterferenceBasedSynchronizationBlockade}
In this section, we discuss how interference effects lead to a novel type of synchronization blockade. 
For clarity of the formulas, we focus here on the equatorial limit cycle but the same quantum effect is present in other oscillators, including the van der Pol limit cycle.

Quantum synchronization blockade was first reported in the study of two coupled anharmonic oscillators, where conservation of energy was found to favor the synchronization of detuned oscillators~\cite{Loerch-PhysRevLett.118.243602}. 
This behavior is in contrast to the classical expectation that synchronization is strongest on resonance. 
However, in the present spin-$1$ system the energy levels are equally spaced, and if there is a synchronization blockade, it has to be of a different physical origin.

\begin{figure}
	\centering
	\includegraphics[width=0.48\textwidth]{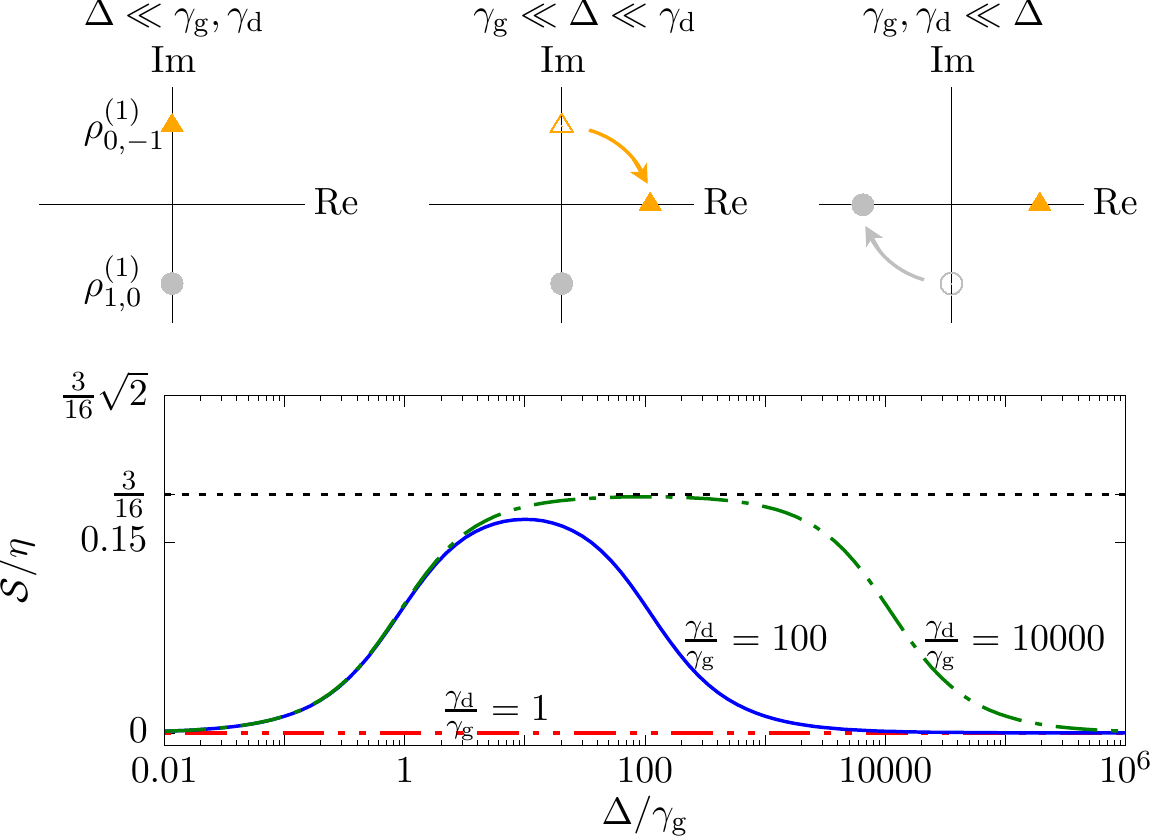}
	\caption{
		Illustration of the interference-based quantum synchronization blockade effect for the equatorial limit cycle introduced in Sec.\,\ref{sec:Methods:CompareDifferentSystems}.
		The relative phase of the signal components is fixed to $\chi = 0$ and their relative amplitude is chosen according to Eq.\,\eqref{eq:optEq}. 
		For imbalanced dissipation rates, $\gamma_\mathrm{d} \neq \gamma_\mathrm{g}$, the coherences $\rho_{0,-1}^{(1)}$ and $\rho_{1,0}^{(1)}$ rotate by different angles if the detuning is in the range $\gamma_\mathrm{g} \lesssim \Delta \lesssim \gamma_\mathrm{d}$, as indicated by the sketches in the upper row.
		Therefore, their destructive interference is partially lifted and synchronization is obtained as shown by the lower plot of $\mathcal{S}/\eta$.
		For strong asymmetries $\gamma_\mathrm{d} \gg \gamma_\mathrm{g}$, the maximum synchronization converges to $\mathcal{S}/\eta = 3/16$, which is indicated by the dotted black line. 
		This is smaller than the maximum synchronization possible for this limit cycle, $3 \sqrt{2}/16$, because the detuning cannot fully align the coherences to interfere constructively. 
		The threshold parameter is $\eta=\ValueFigureVIeta$. 
	}
	\label{fig:6}
\end{figure}

We previously found that for any value of the detuning $\Delta$, there exists a combination of optimal angles~\eqref{eq:optEq} such that the synchronization is maximized~\eqref{eq:maxEq}. 
On resonance, $\Delta=0$, the condition on the relative phase between the tones is $\chi^\mathrm{opt}=\pi$. 
On the other hand, shifting the angle to $\chi=0$ leads to perfect destructive interference $\mathcal{S}/\eta=0$. 
Now if we change the detuning while keeping $\chi = 0$ fixed, the coherences~\eqref{eqn:Equatorial:Coherences} start to rotate clockwise in the complex plane up to an angle of $\pi/2$ for infinitely large detuning. 
This is shown in Fig.~\ref{fig:6}. 
If the dissipation rates are balanced $\gamma_\mathrm{g}=\gamma_\mathrm{d}$, both coherences rotate together and the interference remains destructive regardless of the detuning. 
However, if one of the rates dominates, the rotation of the corresponding coherence lags behind such that the destructive interference is suppressed in a transient regime. 
This is leads to the onset of synchronization away from resonance, as illustrated in the main plot of Fig.\,\ref{fig:6}.

Specifically, the synchronization measure reads
\begin{align}
	\mathcal{S} = \eta \frac{3}{16} \sqrt{1 - \cos\left( \arctan\left[ \frac{(\gamma_\mathrm{d} - \gamma_\mathrm{g}) \Delta}{\gamma_\mathrm{d} \gamma_\mathrm{g} + \Delta^2} \right] \right)} \comma
\end{align}
where the cosine term approaches zero for a strong lag before coming back to unity. 
Maximum synchronization is achieved at $\abs{\Delta}=\sqrt{\gamma_\mathrm{g} \gamma_\mathrm{d}}$, where it converges to $\mathcal{S}/\eta \to 3 /16 \approx 0.188$ in the limit $\gamma_\mathrm{g} \gg \gamma_\mathrm{d}$. 
Note that this value remains below the fundamental limit~\eqref{eq:maxEq} of the equatorial limit cycle, since the detuning is not able to rotate the coherences up to a relative angle of $\pi$, which is the condition for them to interference constructively.

\section{Optimal quantum synchronization}
\label{sec:Bound}
In this section, we derive the fundamental limit to synchronization deep in the quantum regime. 
In contrast to the previous sections, we do not focus on any specific limit cycle. 
Instead we only rely on the properties of the spin-$1$ system supporting the limit cycle, which follow from the laws of quantum mechanics and the paradigm of synchronization. 
This is the first time, to our knowledge, that such an optimization is performed over all signals and all possible limit cycles of a given system.

\subsection{Upper bound for a spin-$1$ system}
In a first step, we derive an upper bound on the synchronization measure $\mathcal{S}(\hat{\rho})$ based on the analytical insights gathered in the previous sections. 
As discussed in Sec.\,\ref{sec:Methods:LC}, the rotational invariance of the limit-cycle state requires a diagonal steady-state density matrix, which we parametrize by
\begin{align}
	\hat{\rho}^{(0)} = \begin{pmatrix}
		\frac{1 - a - \delta}{2} & & \\ & a & \\ & & \frac{1 - a + \delta}{2} 
	\end{pmatrix} \fullstop
	\label{eqn:Bound:LCStructure}
\end{align}
Here $0\leq a\leq 1$ is the population of the equatorial state $\ket{0}$, and $\delta$ is a real parameter that satisfies the conditions $\abs{a \pm \delta} \leq 1$ and characterizes the asymmetry in the populations of the extremal states $\ket{\pm 1}$.

In parallel, in Eq.\,\eqref{eqn:Methods:SynchronizationMeasure} we have identified the coherences between energy eigenstates as the resource of quantum synchronization. 
In the optimal situation where the coherences $\rho_{0,1}$ and $\rho_{0,-1}$ interfere constructively, the first-order correction of the expansion~\eqref{eqn:PT:SynchronizationExpansion} can be parametrized as
\begin{align}
	\hat{\rho}^{(1)} = \begin{pmatrix}
		0 & b & c \\
		b^* & 0 & b \\
		c^* & b^* & 0
	\end{pmatrix} \comma
	\label{eqn:Bound:CohStructure}
\end{align}
where $b$ and $c$ are arbitrary complex parameters~\cite{footnote2}. 
As usual, we further set the phase of $c$ such that the maxima of the $\cos(\phi)$ and $\cos(2 \phi)$ terms in $S(\phi \vert \hat{\rho})$ coincide.

\begin{figure}
	\includegraphics[width=.48\textwidth]{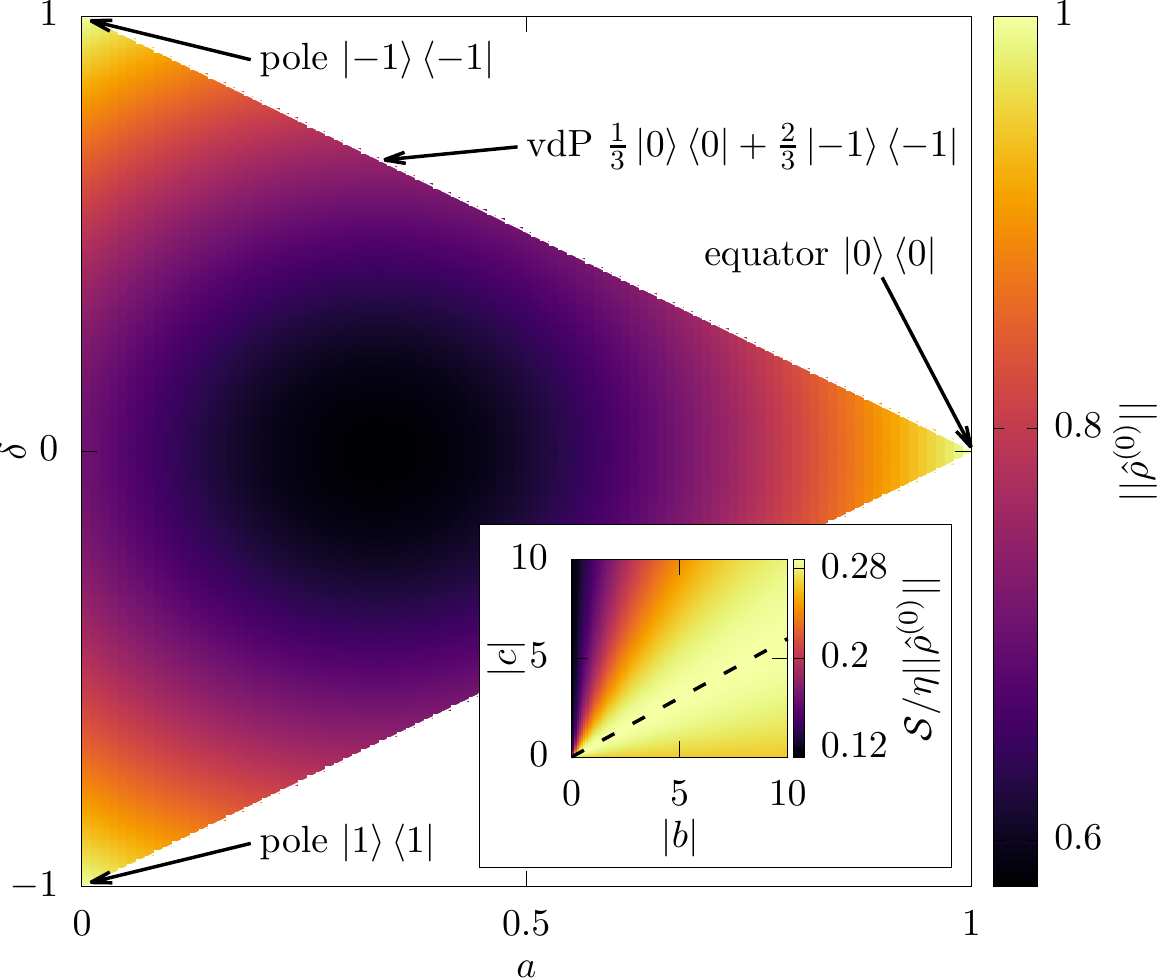}
	\caption{ 
		Value of $\norm{\hat{\rho}^{(0)}}$ for the triangular set of physical limit-cycle states $\hat{\rho}^{(0)}$ parametrized by Eq.\,\eqref{eqn:Bound:LCStructure}. 
		The minimum value of the norm, $1/\sqrt{3}$, is obtained for $(a,\delta) = (1/3,0)$. 
		The maximum value, $1$, is obtained for the extremal points of the triangle, which represent pure states. 
		The van der Pol limit cycle corresponds to the point $(a,\delta) = (1/3,2/3)$. 
		Inset:  
		Factor $(3 \abs{2 b}/8 \sqrt{2} + \abs{c}/2 \pi)/\norm{\hat{\rho}^{(1)}}$ as a function of the coherences $b$ and $c$ introduced in Eq.\,\eqref{eqn:Bound:CohStructure}. 
		The maximum value, $\sqrt{8 + 9 \pi^2/2}/8 \pi$, is achieved along the dashed black line $\abs{b}/\abs{c} = 3 \pi/4 \sqrt{2}$. 
		If the coherence $\rho_{-1,1}$ cannot be built up, $\abs{c}=0$, a value of $3/8 \sqrt{2}$ is obtained. 
	}
	\label{fig:6:PrefactorS}
\end{figure}

Substituting Eq.\,\eqref{eqn:Methods:DefinitionEpsilon} into Eq.\,\eqref{eqn:Methods:SynchronizationMeasureMax} we find that the synchronization measure $\mathcal{S}(\hat{\rho})$ is a product of the term $\eta \norm{\hat{\rho}^{(0)}}$, which depends only on the structure~\eqref{eqn:Bound:LCStructure} of the limit cycle and is shown in the main plot of Fig.\,\ref{fig:6:PrefactorS}, and of the term $(3 \abs{2 b}/8 \sqrt{2} + \abs{c}/2 \pi)/\norm{\hat{\rho}^{(1)}}$, which depends only on the coherences introduced in Eq.\,\eqref{eqn:Bound:CohStructure} and is shown in the inset.
An upper bound to the synchronization achievable in the spin-$1$ system can thus be derived by maximizing both terms individually. 
Specifically, the term $\norm{\hat{\rho}^{(0)}}$ takes its maximal value of unity for any pure state, which are represented by the extremal points of the set of physical states shown in Fig.\,\ref{fig:6:PrefactorS}. 
On the other hand, the second term of $\mathcal{S}(\hat{\rho})$ is maximized when the ratio of the coherences satisfies $\abs{b}/\abs{c} = 3 \pi/(4 \sqrt{2})$, which is indicated by the dashed black line in the inset of Fig.\,\ref{fig:6:PrefactorS}. 
Taking the product of the two maxima, we find that the synchronization measure is bounded from above by 
\begin{align}
	\mathcal{S} \leq \mathcal{S}_\mathrm{max} = \eta \frac{\sqrt{2 (16 + 9 \pi^2)}}{16 \pi} \approx 0.288 \eta \fullstop
	\label{eq:bound}
\end{align}
This result depends on the phase space via the prefactors of the $\cos(\phi)$ and $\cos(2 \phi)$ terms in Eq.~\eqref{eqn:Methods:SynchronizationMeasure}. 
The corresponding number for the phase-space of a harmonic-oscillator introduced in Eq.~\eqref{eqn:vdP:SyncMeasure} is
\begin{align}
	\mathcal{S} \leq \mathcal{S}_\mathrm{max}^\mathrm{osc} = \eta \frac{\sqrt{3}}{2 \sqrt{2} \pi} \approx 0.195 \eta \fullstop
\end{align}

\subsection{Tightness of the bound}
As summarized in Table~\ref{tbl:Summary}, all the combinations of limit cycles and signals considered up to now stay below the bound~\eqref{eq:bound}. 
Therefore, it remains to determine whether any physical limit-cycle oscillator can actually reach the bound $\mathcal{S}_\mathrm{max}$.

\begin{table}
\caption{
	Synchronization performance $\mathcal{S}(\hat{\rho})/\eta$ of the quantum van der Pol and the equatorial limit cycles for different signals. 
	The results are bounded by the maximum synchronization that can be achieved in a spin-$1$ system, $\mathcal{S}_\mathrm{max} = 0.288 \eta$. 
}
\label{tbl:Summary}
\begin{tabular}{l|c|c|cl}
	\hline 
	\hline 
	\multicolumn{1}{c|}{limit cycle} & \multicolumn{3}{c}{signal} \\
	& semi-classical & semi-classical \& squeezing & optimal \\
	\hline
	van der Pol & $0.140$ & $0.163$  & $0.215$ \\
	equatorial   & $0.188$ & $0.188$ & $0.265$ \\
	\hline
	\hline
\end{tabular}
\end{table}

This search is complicated by the trade-off that exists between maximizing $\norm{\hat{\rho}^{(0)}}$ and reaching the optimal ratio $\abs{b}/\abs{c}$. 
To illustrate this point, we can classify the limit cycles studied in the previous sections with respect to these two quantities. 
The van der Pol limit cycle with the optimized signal discussed in Sec.\,\ref{sec:vdpOpt} successfully implements the optimal ratio of the coherences, but, since its limit cycle is a statistical mixture of different spin states, it does not maximize $\norm{\hat{\rho}^{(0)}}$.
On the other hand, the equatorial limit cycle discussed in Sec.\,\ref{eq:equaOpt} implements the optimal value $\norm{\hat{\rho}^{(0)}} = 1$ by stabilizing the pure equatorial state $\ket{0}$, but the symmetry $\rho_{1,1} = \rho_{-1,-1} = 0$ then enforces $\abs{c} = 0$, putting the optimal ratio of the coherences out of reach.

To design a combination of limit cycle and signal that reaches $\mathcal{S}_\mathrm{max}$, we thus need to break the symmetry between the states $\ket{\pm 1}$, while ensuring that the limit cycle remains close to a pure state. 
To this end, we supplement the equatorial limit cycle, $\hat{O}_\mathrm{g} = \hat{S}_+ \hat{S}_z$ and $\hat{O}_\mathrm{d} = \hat{S}_- \hat{S}_z$, by a third decay channel $\hat{O}_\mathrm{d'} = \hat{S}_z \hat{S}_-$ at rate $\gamma_{\mathrm{d}'}$, which induces an asymmetry in the limit cycle, 
\begin{align}
	\hat{\rho}^{(0)} = \begin{pmatrix}
		0 & & \\ 
		& \frac{\gamma_\mathrm{g}}{\gamma_\mathrm{g} + \gamma_\mathrm{d'}} & \\
		& & \frac{\gamma_\mathrm{d'}}{\gamma_\mathrm{g} + \gamma_\mathrm{d'}}
	\end{pmatrix} \fullstop
	\label{eqn:Bound:OptimalLC}
\end{align}
We focus on the regime $\gamma_\mathrm{d'} \ll \gamma_\mathrm{g}$ where the limit cycle remains close to the state $\ket{0}$. 
However, in contrast to the purely equatorial case, the present limit cycle is sensitive to a squeezing signal, \emph{i.e.} we can exploit the small but finite asymmetry in the populations of the extremal states $\ket{\pm 1}$ to engineer a non-vanishing coherence $\abs{c}$. 
In the limit $\gamma_{\mathrm{d}'} \ll \gamma_\mathrm{g}$, the optimal ratio $\abs{b}/\abs{c} = 3 \pi/4 \sqrt{2}$ is obtained by choosing the amplitude
\begin{align}
	\abs{t_{-1,1}} &= \frac{4}{3 \pi} \sqrt{\frac{(\gamma_\mathrm{g} + \gamma_\mathrm{d})^2 + 4 \Delta^2}{\gamma_\mathrm{d}^2 + \gamma_\mathrm{g}^2 + 2 \Delta^2}} \frac{\gamma_\mathrm{g}}{\gamma_{\mathrm{d}'}}  
\end{align}
of the squeezing tone, whereas the angles $\chi^\mathrm{opt}$ and $\zeta^\mathrm{opt}$ are the same as in Eq.\,\eqref{eq:optEq}.
The divergence of the squeezing tone in the limit $\gamma_{\mathrm{d}'} \to 0$, $\abs{t_{-1,1}} \propto \gamma_\mathrm{g}/\gamma_{\mathrm{d}'}$, reflects the fact that the squeezing signal requires an asymmetry between the $\ket{\pm 1}$ states to build up the coherence $\rho_{-1,1}^{(1)}$.
The synchronization measure reads
\begin{align}
	\mathcal{S} = \eta \frac{\sqrt{2 (16 + 9 \pi^2)}}{16 \pi} \sqrt{\frac{\gamma_\mathrm{g}^2 + \gamma_\mathrm{d'}^2}{(\gamma_\mathrm{g} + \gamma_\mathrm{d'})^2}}\underset{\gamma_\mathrm{d'} \ll \gamma_\mathrm{g}}{\longrightarrow}\mathcal{S}_\mathrm{max} \fullstop
\end{align}
Hence, in the regime of interest $\gamma_\mathrm{d'} \ll \gamma_\mathrm{g}$ we find that the synchronization converges to the upper bound $\mathcal{S}_\mathrm{max}$ by approaching the equatorial limit-cycle state with $\norm{\hat{\rho}^{(0)}} \approx 1$ while keeping the ratio of the coherences set to $\abs{b}/\abs{c} = 3 \pi/4 \sqrt{2}$. 
This result demonstrates that the bound~\eqref{eq:bound} is tight and indeed corresponds to the maximum synchronization achievable in the spin-$1$ system.

\section{Discussion}
\label{sec:Discussion}
Quantum synchronization has still not been observed experimentally, despite the existence of proposals with trapped ions~\cite{Lee-PRL.111.234101} and optomechanical~\cite{Walter-PRL.112.094102} oscillators. 
A significant part of the challenge lies in the specific limit cycle that was envisioned at the time, namely the van der Pol oscillator, which requires to engineer a single-photon gain and a damping where photons decay in pairs.

Our findings reveal that one actually has a lot of freedom in tailoring a quantum system that is able to synchronize, opening the realm of possibilities. 
Specifically, the signal and the limit cycle can be significantly modified, with the latter option offering a large and hitherto unexplored choice of both target states and methods to stabilize it without imposing a phase preference.
When aiming for the first observation of quantum synchronization, this freedom can be leveraged to devise the best strategy to accommodate experimental constraints such as the natural relaxation of the system, which is typically considered as an undesired source of noise. 
Shifting the paradigm, we now show that this natural relaxation can in fact be exploited as a useful contribution to the stabilization of the limit cycle, reducing the experimental complexity of implementing a quantum self-sustained oscillator.

\begin{figure}
	\centering	
	\includegraphics[width=.4\textwidth]{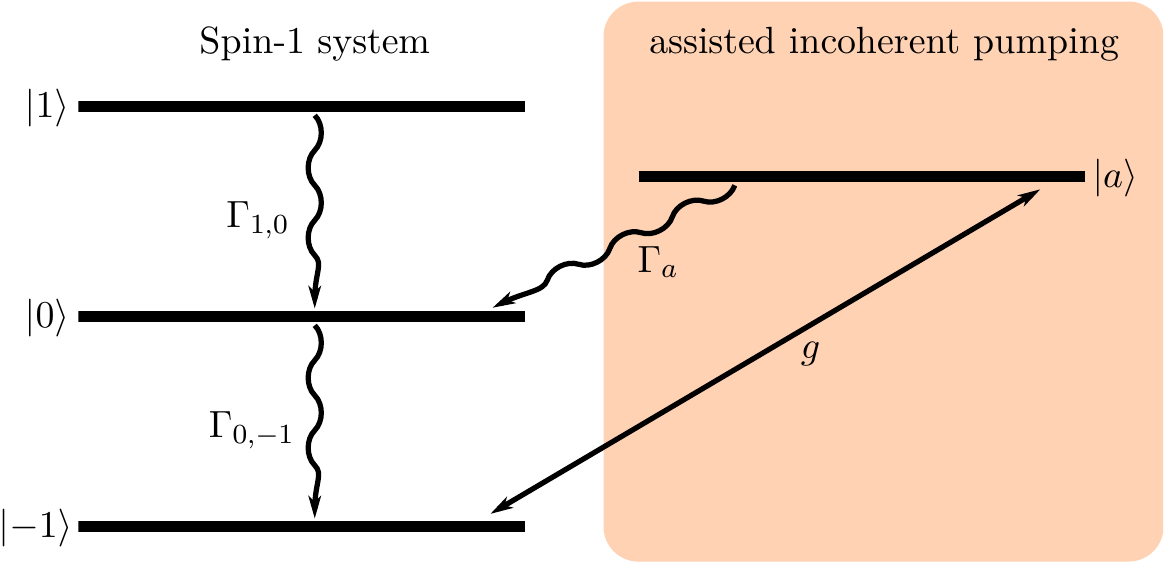}
	\caption{
		Experimental proposal to sustain self-oscillations in a spin-$1$ system. 
		The damping is realized by the natural relaxation of the spin ladder, while the incoherent gain is engineered by coherently driving the $\ket{-1}\leftrightarrow\ket{a}$ transition and exploiting the spontaneous relaxation of the ancilla state $\ket{a}$ to the equatorial state $\ket{0}$.
	}
	\label{fig:9:realization}
\end{figure}

Consider a spin-$1$ system which dissipates energy to its environment at rates $\Gamma_{1,0}$ and $\Gamma_{0,-1}$, as illustrated in Fig.\,\ref{fig:9:realization}. 
This system is realized in a variety of experimental platforms, such as trapped ions \cite{Cohen-PhysRevLett.112.040503,Senko-PhysRevX.5.021026}, nitrogen-vacancy centers \cite{Stark-arxiv.1805.09435}, and superconducting transmons \cite{Neeley-science.325.5941,Bianchetti-PhysRevLett.105.223601}. 
Given that we explicitly include the natural dissipative dynamics into the limit cycle stablization, the only engineering challenge that is left is to stabilize the oscillator away from its ground state by incoherently pumping the transition between the ground state $\ket{-1}$ and the equatorial state $\ket{0}$. 
This is feasible with current technology, and as an example we consider a scheme that has been demonstrated experimentally with superconducting circuits~\cite{Leek-PhysRevB.79.180511,Leek-PhysRevLett.104.100504}. 
There, the working principle is to assist the incoherent transfer from the ground state by driving a transition to an ancilla level, which decays spontaneously into the excited state of interest (see orange box in Fig.\,\ref{fig:9:realization}). 
This technique has been used to efficiently achieve population inversion of up to $93\%$ in the steady-state~\cite{Leek-PhysRevLett.104.100504}. 
Such a pumping scheme, supplemented by the natural relaxation of the system, thus successfully establishes a quantum limit cycle.

We now go beyond the proof-of-concept approach and assess the performance of this minimalistic limit-cycle oscillator, benchmarking against the optimal limit cycle derived in Sec.\,\ref{sec:Bound}. 
In the regime of interest $\Gamma_{0,-1} \ll \Gamma_{a}$ where the population in the ancilla state is negligible, $(\Gamma_{0,-1}/\Gamma_{a})/(1 + 1/4\mathcal{C}) \ll 1$, the steady state of the spin-$1$ system is given by
\begin{align}
	\hat{\rho}^{(0)} = \begin{pmatrix}
		0 & & \\
		& \frac{4 \mathcal{C}}{1 + 4 \mathcal{C}} & \\
		& & \frac{1}{1 + 4 \mathcal{C}}
	\end{pmatrix} \comma
\end{align}
where $\mathcal{C} = g^2/\Gamma_{0,-1} \Gamma_{a}$ denotes the cooperativity of the pumping process. 
The larger the cooperativity, the more efficiently the pumping acts against the natural relaxation. 
In practice, the population of the equatorial state $\ket{0}$ can be varied from zero close to unity by adjusting the cooperativity $\mathcal{C}$: a value of $\mathcal{C} = 1/8$ implements a van der Pol-type occupation distribution, whereas a large cooperativity $\mathcal{C} \gg 1$ implements a limit-cycle state that is mostly the equatorial state $\ket{0}$. 
Remarkably, any finite cooperativity will inevitably lead to an asymmetry between the empty state $\ket{1}$ and the nearly-empty ground state $\ket{-1}$, which is exactly the requirement we derived for optimizing synchronization deep in the quantum regime. 
This implies that the experimental scheme proposed here is actually able to implement the optimal limit cycle provided that the cooperativity is large enough. 
The experimental demonstration of the pumping scheme reported a decade ago~\cite{Leek-PhysRevLett.104.100504} corresponds to $\mathcal{C} \approx 3$. 
This achievement is already large enough to implement the first observation of quantum synchronization, and sets the optimal limit cycle within experimental reach of state-of-the-art platforms.

\section{Conclusion}
\label{sec:Conclusions}
We have developed a framework to study synchronization in the quantum regime based on the perturbative nature of the phenomenon. 
This allowed us to identify the coherences between energy eigenstates as the resource of quantum synchronization. 
Consequently, we have found that interference effects between coherences that transform identically under rotations may either enhance or hinder synchronization. 
This result allowed us to explain previous observations and led us to identify a novel interference-based synchronization blockade that does not rely on an anharmonicity in the energy levels.

Our framework contains a prescription on how to choose the signal strength such that the signal stays within the perturbative regime of synchronization and the integrity of the limit cycle is guaranteed to be preserved. 
The resulting maximum signal strength is a function of the detuning, such that the classic Arnold tongue can be extended for nonzero detuning and becomes a snake-like split tongue.

Focusing on the smallest quantum system that can be synchronized, namely a spin-$1$ system, we have then applied the formalism to compare the synchronization of different combinations of limit cycles and signals. 
To this end, we have first demonstrated that the van der Pol model can be faithfully represented even though the planar position-momentum phase space of the oscillator is replaced by the spherical phase space of a spin. 
Exploiting the low-dimensional Hilbert space, we have been able to provide an analytical description of previous numerical studies and to derive the optimized signal for this specific limit cycle. 
We have then compared the performance to the equatorial limit cycle, which we found to synchronize better despite being insensitive to a squeezing tone.

Finally, the analytical understanding gained along the way led us to derive a fundamental bound on the maximum synchronization that can be achieved in the spin-$1$ system. 
This bound has been shown to be tight by explicitly constructing a limit cycle that reaches the bound asymptotically for an optimized signal. 
Moreover, we have motivated that this limit cycle is actually within experimental reach of current technology by proposing a practical stabilization scheme. 
With this limit-cycle oscillator at hand, quantum synchronization could be readily observed by applying standard coherent (laser) signals that are routinely used in most experimental platforms.

Our findings pave the way to study synchronization of spin-based networks. 
Since the spin-$1$ system has the smallest Hilbert space that is able to capture all features of a van der Pol oscillator deep in the quantum regime, it is a promising candidate to study networks both in terms of numerical efficiency and analytical accessibility. 
Besides, the spin architecture grants access to efficient numerical simulation techniques \cite{Shammah-PRA.98.063815}.

Furthermore, our result on the fundamental limit to the synchronization of a spin $1$ constitutes the first step towards understanding the quantum-to-classical transition. 
It provides a reference point to study how this fundamental limit evolves for higher spin numbers, particularly for half-integer spins which do not have access to an equatorial pure-state limit cycle.

\begin{acknowledgments}
We would like to thank C.\ Bruder and P.\ Magnard for discussions.
This work was financially supported by the Swiss National Science Foundation (SNSF) and the NCCR Quantum Science and Technology.
\end{acknowledgments}

\appendix
\section{Failure of the measure $p_\mathrm{max}(\varepsilon)$}
In this Appendix, we give an example of a limit cycle and a signal for which the deformation measure $p_\mathrm{max}(\varepsilon)$ introduced in Eq.\,\eqref{eqn:Methods:DeformationMeasurePMax} is unable to identify the transition to the forcing regime. 
We consider the van der Pol limit cycle introduced in Sec.~\ref{sec:VdP} of the main text, which is defined by the dissipative coupling operators $\hat{O}_\mathrm{g} = \hat{S}_z \hat{S}_+ - \hat{S}_+ \hat{S}_z/\sqrt{2}$ and $\hat{O}_\mathrm{d} = \hat{S}_-^2/\sqrt{2}$ with the respective rates $\gamma_\mathrm{g}$ and $\gamma_\mathrm{d}$. 
As for the signal we consider the tones $t_{0,1} = r$, $t_{-1,0} = 1/\sqrt{2}$, and $t_{-1,1} = 0$.

\begin{figure}
	\centering
	\includegraphics[width=0.48\textwidth]{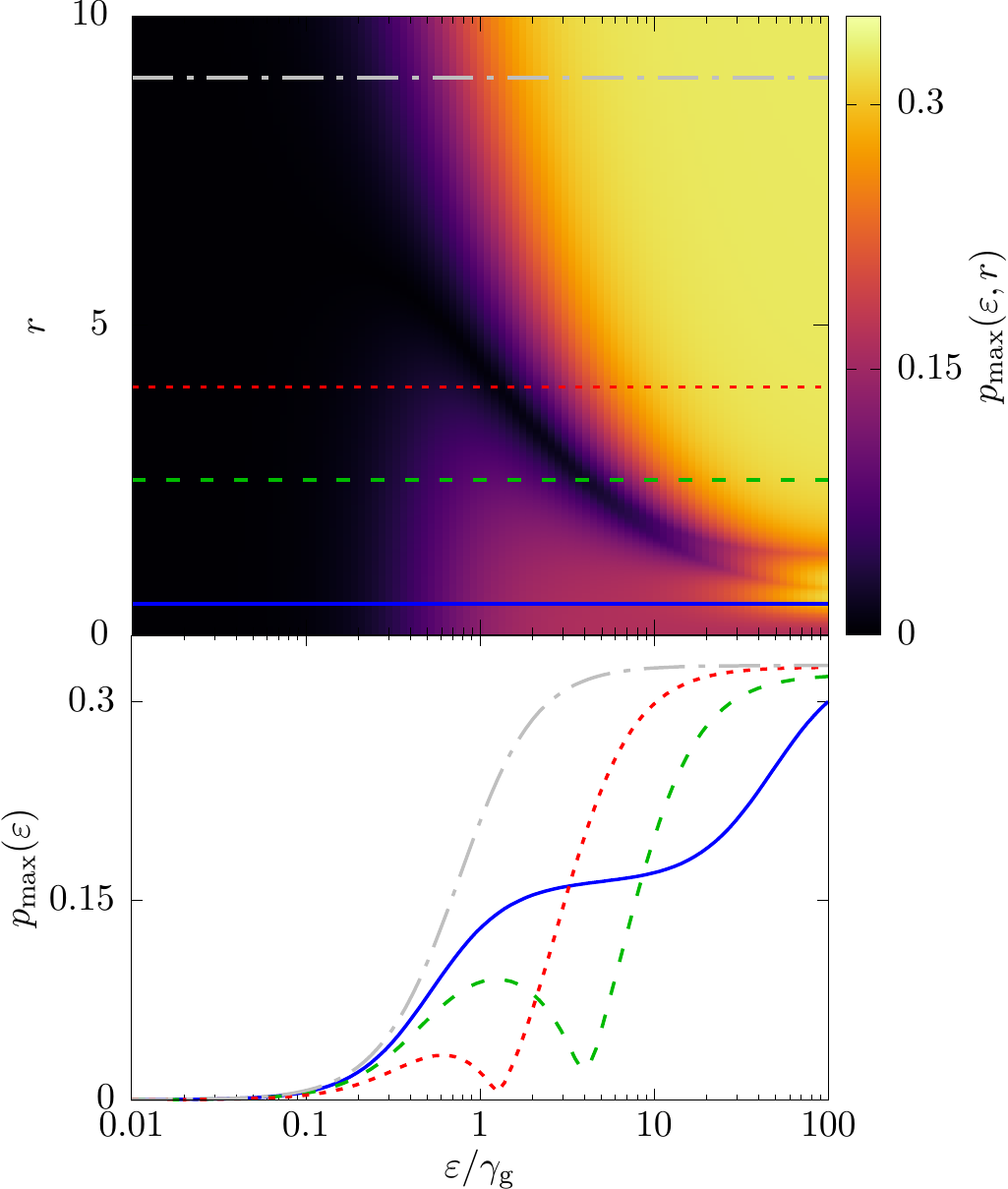}
	\caption{
		Upper panel: Deformation measure $p_\mathrm{max}(\varepsilon,r)$ for a van der Pol limit cycle as a function of the signal strength $\varepsilon$ and the ratio $r = t_{0,1}/\sqrt{2} t_{-1,0}$ of the amplitudes of the semi-classical tones.
		The squeezing tone is switched off, $t_{-1,1} = 0$. 
		Lower panel: Cuts $p_\mathrm{max}(\varepsilon)$ for fixed values $r = \ValueFigureVIIIrA$, $\ValueFigureVIIIrB$, $\ValueFigureVIIIrC$, and $\ValueFigureVIIIrD$, indicated by the corresponding horizontal lines in the upper panel. 
		Parameters are $\gamma_\mathrm{d}/\gamma_\mathrm{g} = \ValueFigureVIIIgammad$ and $\Delta = \ValueFigureVIIIDelta$. 
	}
	\label{fig:Fig8}
\end{figure}

As shown in Fig.\,\ref{fig:Fig8}, there is a range of values $0.6 \lesssim r \lesssim 5.5$ for which the deformation measure $p_\mathrm{max}(\varepsilon)$ is non-monotonous and has a local maximum, then decreases towards 0, before it increases strongly and converges to a constant value in the limit $\varepsilon \to \infty$.
This implies that for a threshold value $\eta$ smaller than the local maximum, there are up to three solutions $\varepsilon_i$ that satisfy $\eta = p_\mathrm{max}(\varepsilon_i)$.

Now for a very pronounced local maximum (cf.\ the dashed green line in Fig.\,\ref{fig:Fig8}), the measure provides a clear indicator that the limit cycle is deformed to an intermediate state for $\varepsilon \gtrsim \min\{ \varepsilon_i \}$, before it converges to another deformed state in the strongly forced regime $\varepsilon \gg \max \{ \varepsilon_i \} $. 
In this situation, $\varepsilon_\mathrm{max} = \min\{ \varepsilon_i \}$ is straightforwardly identified as the maximum signal strength allowed for synchronization. 
However, the value of the local maximum decreases with $r$, and in particular for $r \approx 5$ the peak almost vanishes (cf.\ the dotted red line in Fig.\,\ref{fig:Fig8}). 
This means that for any fixed value of the threshold $\eta$ there is an $r$ such that the first deviation of $p_\mathrm{max}(\varepsilon)$ is not detected, without having a physical argument that it does not belong to the forcing regime. 
Consequently, the measure $p_\mathrm{max}(\varepsilon)$ fails to give a definite answer for the transition to the forcing regime.


\begin{thebibliography}{999}
\bibitem{footnote1}
	Note that we are intentionally omitting diagonal operators $\hat{O}_j$ spanned by $\hat{S}_z$ and $\hat{S}_z^2$, which would in principle be allowed, as their effect is to generate dephasing without contributing to the stabilization of the amplitude, thereby degrading  synchronization.
\bibitem{footnote2}
	For simplicity, we assume the symmetry $\rho_{1,0} = \rho_{0,-1} = b$ which can easily be derived from a more general parametrization of $\hat{\rho}^{(1)}$.
\end{thebibliography}
\end{document}